\documentclass{article}

\def\xxinput#1{\input#1}

\xxinput{vsolj02.sty}

\usepackage{longtable}
\usepackage{graphicx}
\usepackage[comma,colon]{natbib}

\usepackage[OT2,T1]{fontenc}

\def\cite{\citealt}
\setcitestyle{aysep={}}

\def\stageBmath{{\rm (stage\; B)}}
\def\stageCmath{{\rm (stage\; C)}}
\def\math31{{\rm (3\!:\!1)}}

\begin{document}

\title{Evolution of short-period cataclysmic variables: implications from eclipse}
\vskip -2mm
\title{modeling and stage A superhump method (with New Year's gift)}

\begin{figure*}[h]
  \begin{center}
    \includegraphics[width=16cm]{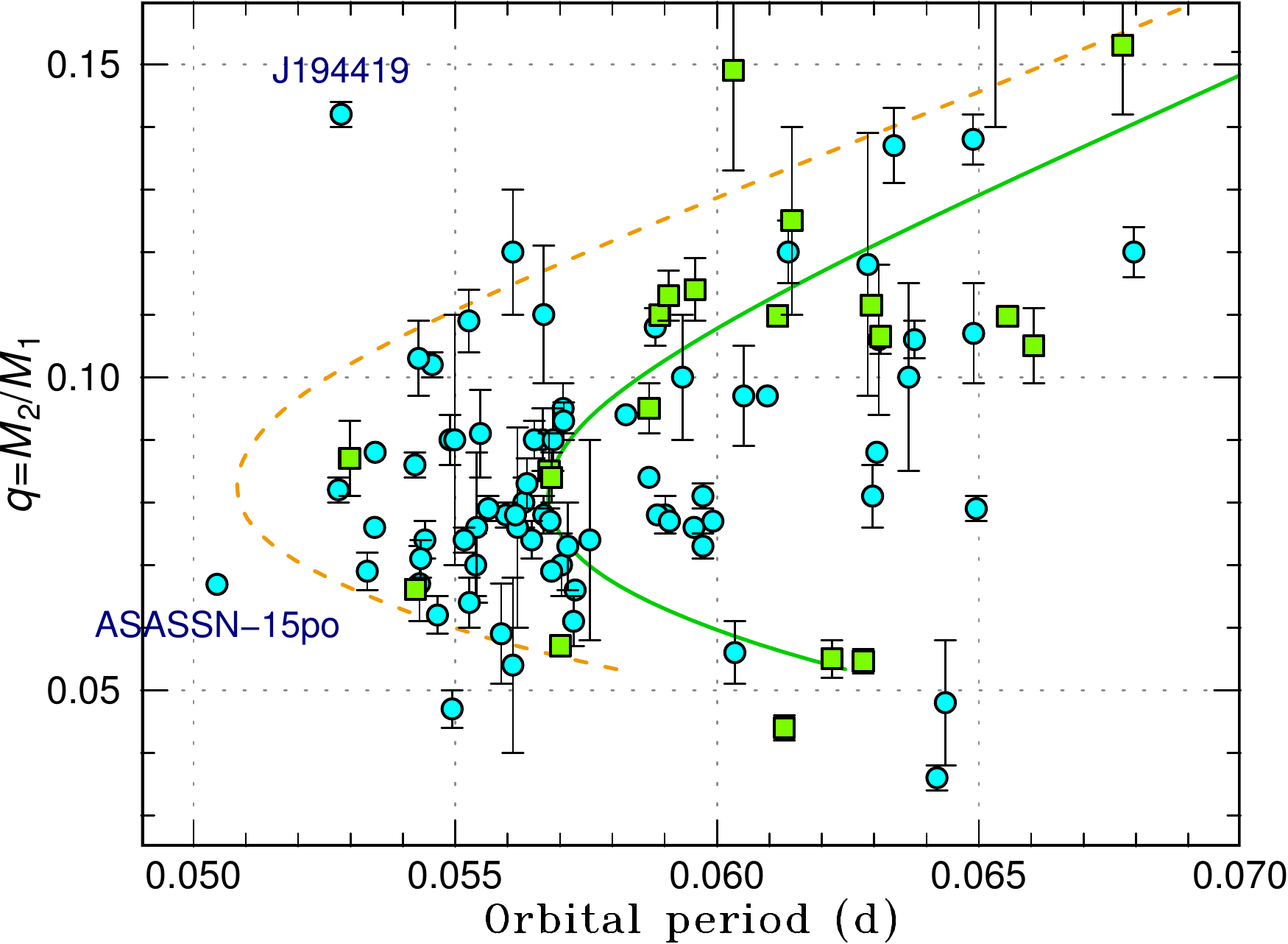}
  \end{center}
  \caption{Mass ratios ($q$) versus orbital periods ($P_{\rm orb}$)
  determined by the eclipse modeling method and the stage A superhump
  method, enlargement around the period minimum.
  See text for the detail; the answer for the symbols is shown
  in figure \ref{fig:qallnew}.
  The dashed and solid curves represent the standard and optimal
  evolutionary tracks in \citet{kni11CVdonor}, respectively.
  }
  \label{fig:qalllarge}
\end{figure*}

\author{Taichi Kato$^1$}
\author{$^1$ Department of Astronomy, Kyoto University,
       Sakyo-ku, Kyoto 606-8502, Japan}
\email{tkato@kusastro.kyoto-u.ac.jp}

   (\textbf{Abstract} is given at the end of the paper).

\section*{Prologue}

   ``Look to the skies, and you will feel it --- a deep
universal fascination ...'' --- stargazers will completely
agree with the phrase, but this is the beginning of
the narration of the TV documentary ``Extraordinary Birds''
directed by Tom Simon in 2000.  I refrain from talking about
my favorite birds for a while;
now, just look at figure \ref{fig:qalllarge}!
Although this may not be as fascinating as what you see
in the skies, this is a figure summarizing our current
best knowledge about the terminal evolution of
cataclysmic variables (CVs).
CVs are close binaries consisting of a white dwarf and
a mass-transferring low-mass dwarf star.
CVs evolve from upper right on this figure from
long orbital periods ($P_{\rm orb}$) to shorter $P_{\rm orb}$
and to lower mass ratios ($q$) by transferring the matter
from the secondary.  CVs then reach the ``period minimum'',
after which $P_{\rm orb}$ lengthens while $q$ is still
decreasing.  Such objects are called ``period bouncers''.

   This figure contains two symbols representing
measurements by two state-of-the-art methods to determine $q$.
One method employs \textbf{3.5--8.2~m telescopes} equipped with
a specially designed \textbf{multicolor high-speed camera}
(there were even Nature and Science papers among them:
\cite{lit06j1035,her16j1433}).
The other employs \textbf{20--50~cm telescopes},
which are often owned by \textbf{amateur astronomers}
equipped with \textbf{off-the-shelf CCD cameras}.
Can you tell which symbol is which?

   This question would be difficult to answer:
these two methods give almost the same results and
mutually reinforce the reliability each other.
In other words, the second method (stage A superhump
method) is as reliable as the first method
(eclipse modeling).  I will explain the reason in
the following sections.
If you are interested to contribute to observations by
the second method, a book ``Cataclysmic Variable Stars:
How and why they vary'' by \citet{hel01book} will be
helpful.  Our Variable Star Network team
(VSNET Collaboration: \cite{VSNET})
regularly receive observations of superhumps from
amateurs and professionals worldwide and your observations
will surely contribute to reveal the secrets of CVs.

   It is a pity, however, that the measurements of $q$
by the stage A superhump method tend to be
neglected by researchers of the CV evolution,
probably due to a persistent misunderstanding that
the reliability of superhumps for determining $q$ is limited
partly because it is dependent on experimental calibration
based on old knowledge before the 2010s.
In this paper, I review the history of
the misunderstanding, the current reliable method
and a comparison with the results of the eclipse modeling method
using a high-speed photometer, which is usually considered
to be most accurate.

\section{Historical Development}

   Superhumps in SU UMa-type dwarf novae have periods
(superhump period, $P_{\rm SH}$) a few percent longer than
the orbital period $P_{\rm orb}$ [for general information
of cataclysmic variables and dwarf novae,
see e.g. \citet{war95book}].
Superhumps are widely accepted to be caused by
a precessing eccentric accretion disk which arises from 
the 3:1 resonance \citep{whi88tidal,hir90SHexcess,lub91SHa}.
The presence of the gravity of the secondary star causes
the deviation of the gravitational field from
the inverse square law by the white dwarf primary
[see e.g. \citet{hir90SHexcess} for a mathematical treatment],
it is natural to consider that the precession rate
($\omega_{\rm pr}$) can be used as a measure of
the binary mass ratio $q=M_2/M_1$, where $M_1$ and $M_2$
are the masses of the primary (white dwarf) and the secondary
which transfers matter to the primary, respectively.
The relation between $P_{\rm orb}$, $P_{\rm SH}$ and
$\omega_{\rm pr}$ is:
\begin{equation}
\label{equ:precessionrate}
\epsilon^* \equiv \omega_{\rm pr}/\omega_{\rm orb}=1-P_{\rm orb}/P_{\rm SH},
\end{equation}
where $\omega_{\rm orb}$ is the orbital angular 
frequency of the binary.
In actual observations, the fractional superhump excess ($\epsilon$)
is widely used:
\begin{equation}
\label{equ:shexcess}
\epsilon \equiv P_{\rm SH}/P_{\rm orb}-1.
\end{equation}
The relation between $\epsilon$ and $\epsilon^*$ is:
\begin{equation}
\label{equ:shexcessrelation}
\epsilon^*=\epsilon/(1+\epsilon).
\end{equation}

   The fractional superhump excesses were, however, historically
only used to derive approximate values of $q$ mainly due to
the two reasons:
\begin{enumerate}
\item[(1)] The precession rate depends on the radius
or the mass distribution of the disk, which is usually difficult
to determine by observations.
\item[(2)] The precession rate usually does not reflect
the purely dynamical precession.  The pressure effect
slows down the precession rate.
\citep{lub92SH,hir93SHperiod,mur98SH,mon01SH,pea06SH},
while the pressure effect is difficult to formulate
(see e.g. \cite{mon01SH,pea06SH})
or measure by observations.
\end{enumerate}

   Before the identification of the nature of superhumps,
\citet{sto84tumen} made a pioneering work illustrating
that there is a linear relation between $\epsilon$
and $P_{\rm orb}$.  This relation was updated by
\citet{rob87swumaQPO}.
This Stolz-Schoembs relation has widely been used
to estimate $P_{\rm orb}$ from $P_{\rm SH}$
(e.g. \cite{RitterCV3}).
This relation also implied that $\epsilon$ should be
a strong function of $q$ following the evolution of
cataclysmic variables \citep{war76CVevol,pat84CVevolution}.
\citet{mol92SHexcess} first systematically studied
the relation between $\epsilon$ and $P_{\rm orb}$,
and then $q$.  They found very good linear relations
between $\epsilon$ and $P_{\rm orb}$, and
between $\epsilon$ and $q$ assuming main-sequence
secondaries.  This Molnar-Kobulnicky relation was used to
estimate $q$ from $P_{\rm SH}$ or $\epsilon$
(such as \cite{how93efpeg,kat01hvvir}).
\citet{lem93tleo,ski93bklyn,lei94hvvir}
presented updated figures of the Molnar-Kobulnicky
relation.  The work by \citet{min92BHXNSH} was
one of the first to directly estimate $q$ from
$\epsilon$.  They introduced $\eta$, the ratio between
the disk radius and the radius of the 3:1 resonance,
and estimated $\eta$ from measurements of SU UMa stars
and applied to superhumps in black-hole X-ray transients
to estimate the masses of the black holes.
This empirical calibration, however, was not widely
used by researchers of cataclysmic variables
[I could only find \citet{ret97v1974cygSH}].

\begin{figure*}
  \begin{center}
    \includegraphics[height=11cm]{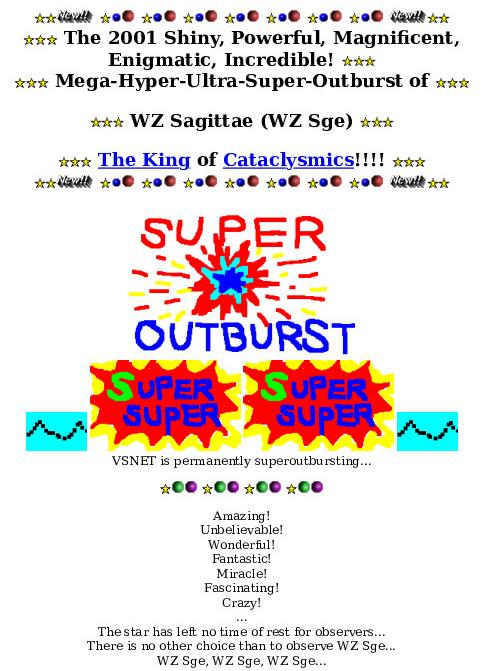}
    \includegraphics[height=11cm]{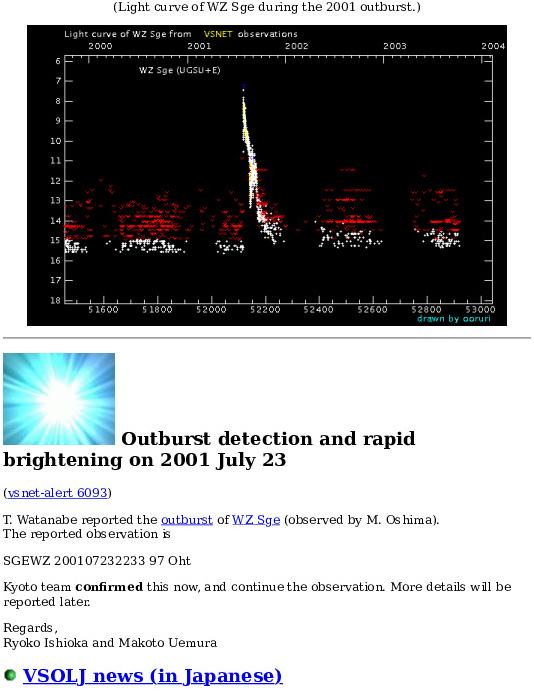}
  \end{center}
  \caption{VSNET page telling the excitement of the unexpected
     outburst of WZ Sge in 2001.}
  \label{fig:wzsge2001}
\end{figure*}

   An independent effort to calibrate the $\epsilon$-$q$
relation became necessary following an burst of new detections
of superhumps in WZ Sge stars
[the examples being \textbf{HV Vir in 1992}: \citet{lei94hvvir,kat01hvvir},
\textbf{AL Com in 1995}: \citet{pyc95alcom,kat96alcom,how96alcom,pat96alcom,nog97alcom},
\textbf{EG Cnc in 1996}: \citet{pat98egcnc,kat04egcnc},
\textbf{V2176 Cyg in 1997}: \citet{nov01v2176cyg,kwa98v2176cyg},
\textbf{V592 Her in 1998}: \citet{due98v592her,kat02v592her},
\textbf{WZ Sge in 2001}: \citet{pat02wzsge,ish02wzsgeletter,bab02wzsgeletter},
figure \ref{fig:wzsge2001};
see \citet{kat15wzsge} for a modern review of WZ Sge stars].
These WZ Sge stars drew attention of researchers since
they have very low-mass secondaries which may be brown dwarfs.
\citet{pat98evolution} derived an equation
\begin{equation}
\epsilon = \frac{0.23 q}{1+0.27 q}
\end{equation}
combined with the analytical formula of the precession rate.
\citet{pat01SH} further calibrated the relation using observed
superhump excesses and obtained
\begin{equation}
\epsilon = 0.216(\pm 0.018)q.
\end{equation}
This relation was derived from various classes of
objects ranging from dwarf novae to novalike variables
and X-ray binaries.  \citet{pat05SH} published
a refinement of the empirical relation
\begin{equation}
\label{equ:pat05eq}
\epsilon = 0.18 q + 0.29 q^2
\end{equation}
based on a broader sample of objects ranging from
dwarf novae to novalike variables.  In \citet{pat01SH},
only one X-ray binary (KV UMa) was used in contrast to
\citet{pat01SH}.
These three formulae are still most widely used.
These ``Patterson'' relations, however, have intrinsic
difficulties.  In addition to the reasons listed earlier
in this section, there are difficulties:
\begin{enumerate}
\item[(3)] The pressure effect is expected to be
different depending on the state of the accretion disk
(see e.g. \cite{pea06SH}).  Patterson's calibration
relied on different states:
non-stationary outbursting dwarf novae,
steady state novalike disks and an X-ray binaries,
which can have disk sizes and temperatures different
from CVs and contributions of the pressure effect
may be different.
\item[(4)] Patterson's formulae assumed $\epsilon$=0
at $q$=0, which is incorrect if the pressure effect
is taken into account.
\end{enumerate}

   According to \citet{lub92SH}, the (apsidal) precession rate
can be written as a form:
\begin{equation}
\omega_{\rm pr}=\omega_{\rm dyn}+\omega_{\rm pressure}+\omega_{\rm stress},
\label{equ:Lubows}
\end{equation}
where the first term, $\omega_{\rm dyn}$, represents a contribution 
to disk precession due to the gravitational potential of
the secondary, giving rise to prograde precession,
the second term, $\omega_{\rm pressure}$ (negative value),
the pressure effect giving rise to retrograde precession, and 
the last term, $\omega_{\rm stress}$, the minor wave-wave interaction.
The functional form of $\omega_{\rm dyn}$ is given
in the next section.  As one could naturally see,
$\omega_{\rm dyn}$ is small when $q$ is small.
$\omega_{\rm pressure}$ is, however, not very dependent on
$q$ and the effect of this term can become larger than
that of the first term for very small $q$.  This is the reason
why the item (4) is important.  The same issue was also
pointed out by \citet{goo06SH}.

\section{Modern Method Using Stage A Superhumps}\label{sec:stageAmethod}

\subsection{Superhumps and dynamical precession}

   The situation has changed since the identification
of superhump stages (stages A, B and C: \cite{Pdot}).
\citet{Pdot} showed that the periods of superhumps
systematically vary.  Stage A superhumps appear first
when superhumps grow.
It was not certain at that time which stage is suitable
for estimating $q$.  In \citet{kat13qfromstageA},
however, stage A superhumps were identified to reflect
the dynamical precession rate of the disk at the radius
of the 3:1 resonance.  Stage B superhumps with smaller
$\epsilon$ is most strongly affected by the pressure effect
and was found to be inadequate to derive $\epsilon$
[Patterson's formulae used stage B superhumps
for dwarf novae; see footnote 11 in \citet{pat11CVdistance}].
Following the treatment in \citet{kat13qfromstageA},\footnote{
   The equations in this part are not absolutely necessary to
   understand the stage A superhump method and its applications.
   One can skip this part and proceed to the next section
   if necessary.
}
\begin{equation}
\label{equ:precession}
\frac{\omega_{\rm dyn}}{\omega_{\rm orb}} = Q(q) R(r),
\end{equation}
$r$ is the dimensionless radius measured in units of the binary 
separation $A$.  The dependence on $q$ and $r$ can be
described as (cf. \cite{hir90SHexcess})
\begin{equation}
\label{equ:qpart}
Q(q) = \frac{1}{2}\frac{q}{\sqrt{1+q}},
\end{equation}
and
\begin{equation}
\label{equ:rpart}
R(r) = \frac{1}{2}\sqrt{r} b_{3/2}^{(1)}(r),
\end{equation}
where
$\frac{1}{2}b_{s/2}^{(j)}$ is the Laplace coefficient\footnote{
   Please don't be discouraged by a formula with an integral.
   Modern computer languages have functions for numerical
   integrations and this integral can be very quickly computed.
}
\begin{equation}
\label{equ:laplace}
\frac{1}{2}b_{s/2}^{(j)}(r)=\frac{1}{2\pi}\int_0^{2\pi}\frac{\cos(j\phi)d\phi}
{(1+r^2-2r\cos\phi)^{s/2}}.
\end{equation}
There is also a polynomial expression \citep{pea03amcvnSH,pea06SH}:
\begin{equation}
\label{equ:precesspoly}
\frac{\omega_{\rm dyn}}{\omega_{\rm orb}}
= \frac{3}{4}\frac{q}{\sqrt{1+q}}r^{3/2}\sum_{n=1}^\infty c_n r^{2(n-1)},
\end{equation}
where
\begin{equation}
\label{equ:laplacepoly}
c_n = \frac{2}{3}(2n)(2n+1)\prod_{m=1}^n~\left({\frac{2m-1}{2m}}\right)^2.
\end{equation}
The full polynomial formula for the dynamical precession rate can be
written down as
\begin{equation}
\label{equ:precesspoly}
\frac{\omega_{\rm dyn}}{\omega_{\rm orb}}
= \frac{3}{4}\frac{q}{\sqrt{1+q}}r^{3/2}\left(1+\frac{15}{8}r^2+\frac{175}{64}r^4+\frac{3675}{1024}r^6+\frac{72765}{16384}r^8+\frac{693693}{131072}r^{10}+\frac{6441435}{1048576}r^{12}+...\right).
\end{equation}
The convergence of this formula is rather slow and it requires
12--13 terms (i.e. up to $r^{22}$ or $r^{24}$) to obtain
a precision of 10$^{-6}$ around the radius the 3:1 resonance.

\subsection{Stage A superhumps and mass ratio}

   During stage A, it is considered that the superhump wave
is confined to the 3:1 resonance region and
the the pressure effect can be neglected
[see figure 13 in \citet{osa13v344lyrv1504cyg} or
figure 4 in \citet{nii21asassn18ey} for schematic
representations of the pressure effect and the regions
of the superhump wave].
One can substitute $r$ by the radius of the 3:1 resonance.
\begin{equation}
\label{equ:radius31}
r_{3:1}=3^{(-2/3)}(1+q)^{-1/3},
\end{equation}
Then $Q(q) R(r_{3:1})$ becomes a function of $q$ and
we can directly estimate $q$ from $\epsilon^*$ of
stage A superhumps.

   Originally in \citet{kat13qfromstageA}, the equations
(\ref{equ:precession}) to (\ref{equ:rpart})
were combined and described as
\begin{equation}
\label{equ:presfreqold}
\frac{\omega_{\rm dyn}}{\omega_{\rm orb}} = \frac{q}{\sqrt{1+q}}\Bigl[\frac{1}{4}\frac{1}{\sqrt{r}}\frac{d}{dr}\Bigr(r^2\frac{db_{1/2}^{(0)}}{dr}\Bigr)\Bigr]
= \frac{q}{\sqrt{1+q}} \Bigl[\frac{1}{4}\frac{1}{\sqrt{r}} b_{3/2}^{(1)}\Bigr] \quad \Leftarrow {\rm incorrect!}
\end{equation}
There was a typo introduced while writing down
an equation in \LaTeX \,in equation (\ref{equ:presfreqold}) =
equation (1) in \citet{kat13qfromstageA},
and the correction was made in \citet{kat16rzlmi}.
The correct equation is
\begin{equation}
\label{equ:presfreqnew}
\frac{\omega_{\rm dyn}}{\omega_{\rm orb}} = \frac{q}{\sqrt{1+q}}\Bigl[\frac{1}{4}\frac{1}{\sqrt{r}}\frac{d}{dr}\Bigr(r^2\frac{db_{1/2}^{(0)}}{dr}\Bigr)\Bigr]
= \frac{q}{\sqrt{1+q}} \Bigl[\frac{1}{4} \sqrt{r} b_{3/2}^{(1)}\Bigr] \quad \Leftarrow {\rm correct}.
\end{equation}
The same incorrect equation was written in
\citet{kat13j1222}, \citet{nak13j2112j2037} and \citet{kat15wzsge}.
The figure and table dealing with this equation
in \citet{kat13qfromstageA} were correct.  No published
$q$ values by the stage A superhump method were affected
by this typo.

\begin{figure*}
  \begin{center}
    \includegraphics[width=16cm]{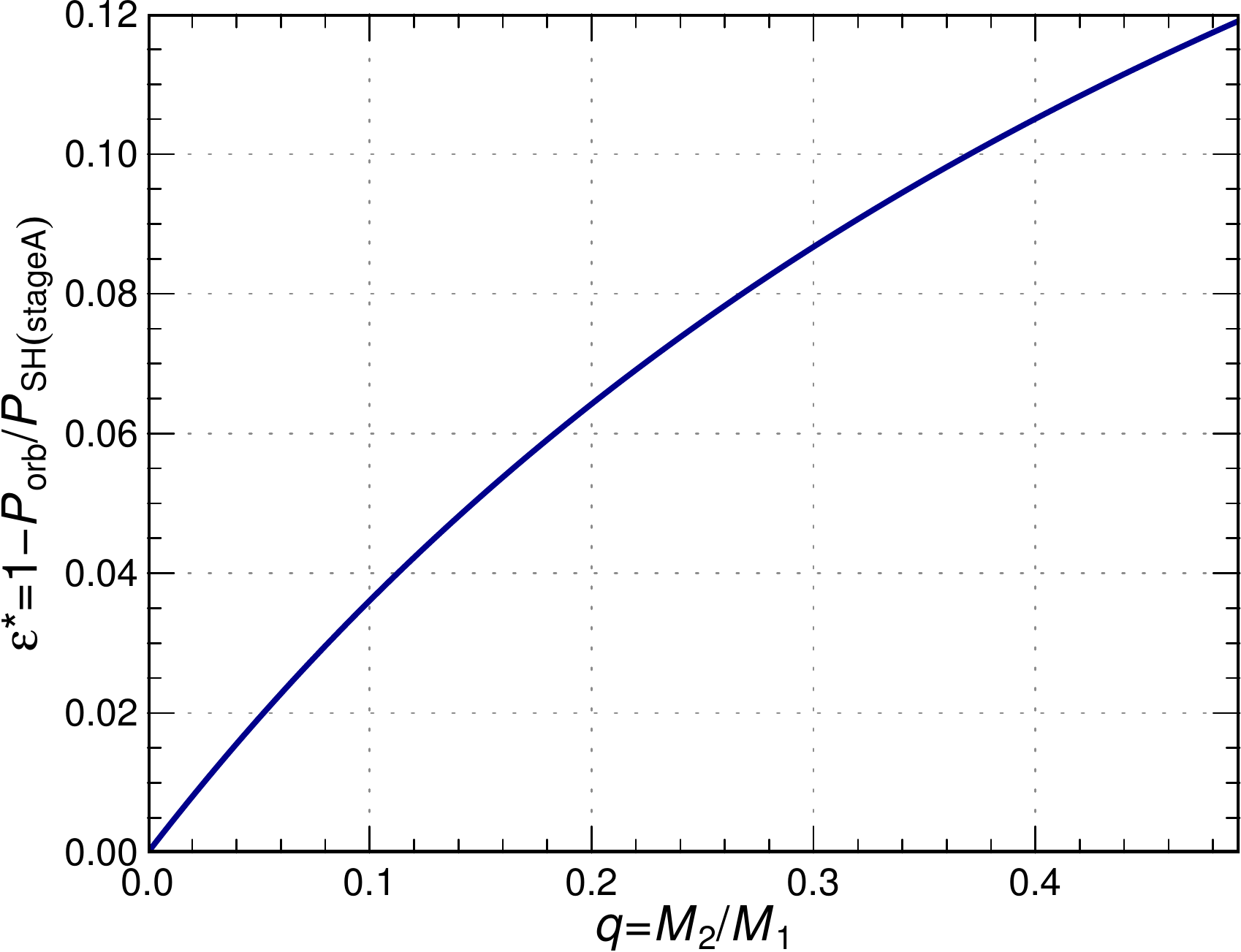}
  \end{center}
  \caption{Relation between $\epsilon^*$ of stage A superhumps and $q$.}
  \label{fig:qeps31}
\end{figure*}

   The stage A superhump method is a \textbf{dynamical}
method to determine $q$ in that it relies only on
celestial mechanics.  The equation is \textbf{analytical}
and \textbf{no experimental calibration is needed}.
These features are clearly advantageous over
the classical superhump methods such as
the Patterson relations.  For users' convenience for interpolation,
I provide an extended version of table 1 and figure 2
in \citet{kat13qfromstageA} in table \ref{tab:stageaepsq} and
figure \ref{fig:qeps31}.
Note that the values are given for (probably) unrealistic
values of $\epsilon^*$.
There is also a polynomial expression of this relation
[equation (7) in \citet{war95suuma} = equation (3.41b) in \citet{war95book}]:
\begin{equation}
\label{equ:presfreqwarner}
\frac{\omega_{\rm dyn}}{\omega_{\rm orb}}\bigg|_{r=r_{3:1}}
= \frac{1}{4}\frac{q}{1+q}\biggl[1+\frac{0.433}{(1+q)^{2/3}}
+\frac{0.146}{(1+q)^{4/3}}+\frac{0.044}{(1+q)^2}+\frac{0.013}{(1+q)^{8/3}}+...\biggr].
\end{equation}
I have confirmed that this equation (up to this term) gives
the same value in table \ref{tab:stageaepsq} to a precision
of 0.001 in $\epsilon^*$.
Although there is also a polynomial equation (4) in
\citet{kat13qfromstageA}, please do not cite this equation
since it is a \textbf{regression}, not an analytical formula.

\begin{table*}
\caption{Relation between $\epsilon^*$ of stage A superhumps and $q$.}
\label{tab:stageaepsq}
\begin{center}
\begin{tabular}{cc|cc|cc|cc|cc}
\hline\hline
$\epsilon^*$ & $q$ & $\epsilon^*$ & $q$ & $\epsilon^*$ & $q$ & $\epsilon^*$ & $q$ & $\epsilon^*$ & $q$ \\ 
\hline
\xxinput{epsout.inc}
\hline
\end{tabular}
\end{center}
\end{table*}

   This stage A superhump method has been applied to many
SU UMa and WZ Sge stars, such as in a series of papers
\citet{Pdot5}--\citet{Pdot10}.
At the time of writing of the present paper, $q$ values have been
determined by the stage A superhump method for more than
100 objects.
There have been indirect estimations of $q$ values using
\citet{kat13qfromstageA}, such as a combination of
the periods of stage A superhumps and post-superoutburst
superhumps \citep{kat13j1222,kat19nyser}.

\subsection{Relation between stage B superhump period
and mass ratio}\label{sec:stagebtoq}

   Using the empirical relation between fractional superhump
excesses of stage A and stage B superhumps
[equation (9) in \cite{kat13qfromstageA}]:
\begin{equation}
\label{equ:stageabrelation}
\epsilon^*\math31=0.016(3)+0.94(12)\epsilon\stageBmath,
\end{equation}
where $\epsilon^*\math31$ is the expected precession rate
at the 3:1 resonance.
This indirect method is helpful when the periods of
stage A superhumps are unknown but uncertainties
resulting from an experimental calibration remain
(the problem of $\epsilon$=0 at $q$=0 in 
Patterson's formulae is, however, avoided).
Examples of applications of this method can be found in
\citet{pav21j1727,shu21aylac}.
These results are not included in the analysis in this paper.

   In this paper, I provide a new calibration since
the number of calibrators has dramatically increased since
\citet{kat13j1222}.  Most of them are from a series of
papers \citet{Pdot}--\citet{Pdot10} (hereafter ``Pdot'' papers;
see section \ref{sec:data} for the detail).
I also used $\epsilon^*$ instead of
$\epsilon$ since the former reflects the precession rate.
The calibrators are given in tables \ref{tab:stagebtab}
(I removed one doubtful measurement of stage B superhumps
ASASSN-16os) for the $q$ values obtained by the stage A superhump
method and table \ref{tab:eclqshbtotab} the $q$ values obtained
by the eclipse modeling method.
The $P_{\rm SH}$ values in these tables refer to the mean
period during stage B.
When there were several measurements of superhumps,
I selected the best one (longest baseline or smallest error).
I sometimes averaged two measurements when the quality of
the two were comparable (see individual reference for details
of the selection of the data).

\begin{center}
\begin{longtable}{lcccc}
\caption{Calibrators for $\epsilon^*$(stage B)-$q$ relation.  The $q$ values
were measured by the stage A superhump method.}\label{tab:stagebtab} \\
\hline\hline
Object & $P_{\rm orb}$ (d) & $q$ (stage A) & $P_{\rm SH}$ (stage B) & $P_{\rm SH}$ reference \\
\hline
\endfirsthead
\caption{Calibrators for $\epsilon^*$(stage B)-$q$ (by stage A superhump method) relation (continued).} \\
\hline\hline
Object & $P_{\rm orb}$ (d) & $q$ (stage A) & $P_{\rm SH}$ (stage B) & $P_{\rm SH}$ reference \\
\hline
\endhead
\hline
\endfoot
\endlastfoot
\xxinput{stagebtab.inc}
\hline
\end{longtable}
\end{center}

\begin{center}
\begin{longtable}{lccccc}
\caption{Calibrators for $\epsilon^*$(stage B)-$q$ relation.  The $q$ values
were measured by the eclipse modeling method.}\label{tab:eclqshbtotab} \\
\hline\hline
Object & $P_{\rm orb}$ (d) & $q$ (eclipse) & Error & $P_{\rm SH}$ (stage B) & $P_{\rm SH}$ reference \\
\hline
\endfirsthead
\caption{Calibrators for $\epsilon^*$(stage B)-$q$ (by eclipse method) relation (continued).} \\
\hline\hline
Object & $P_{\rm orb}$ (d) & $q$ (eclipse) & Error & $P_{\rm SH}$ (stage B) & $P_{\rm SH}$ reference \\
\hline
\endhead
\hline
\endfoot
\endlastfoot
\xxinput{eclqshbtotab.inc}
\hline
\end{longtable}
\end{center}

\begin{figure*}
  \begin{center}
    \includegraphics[width=16cm]{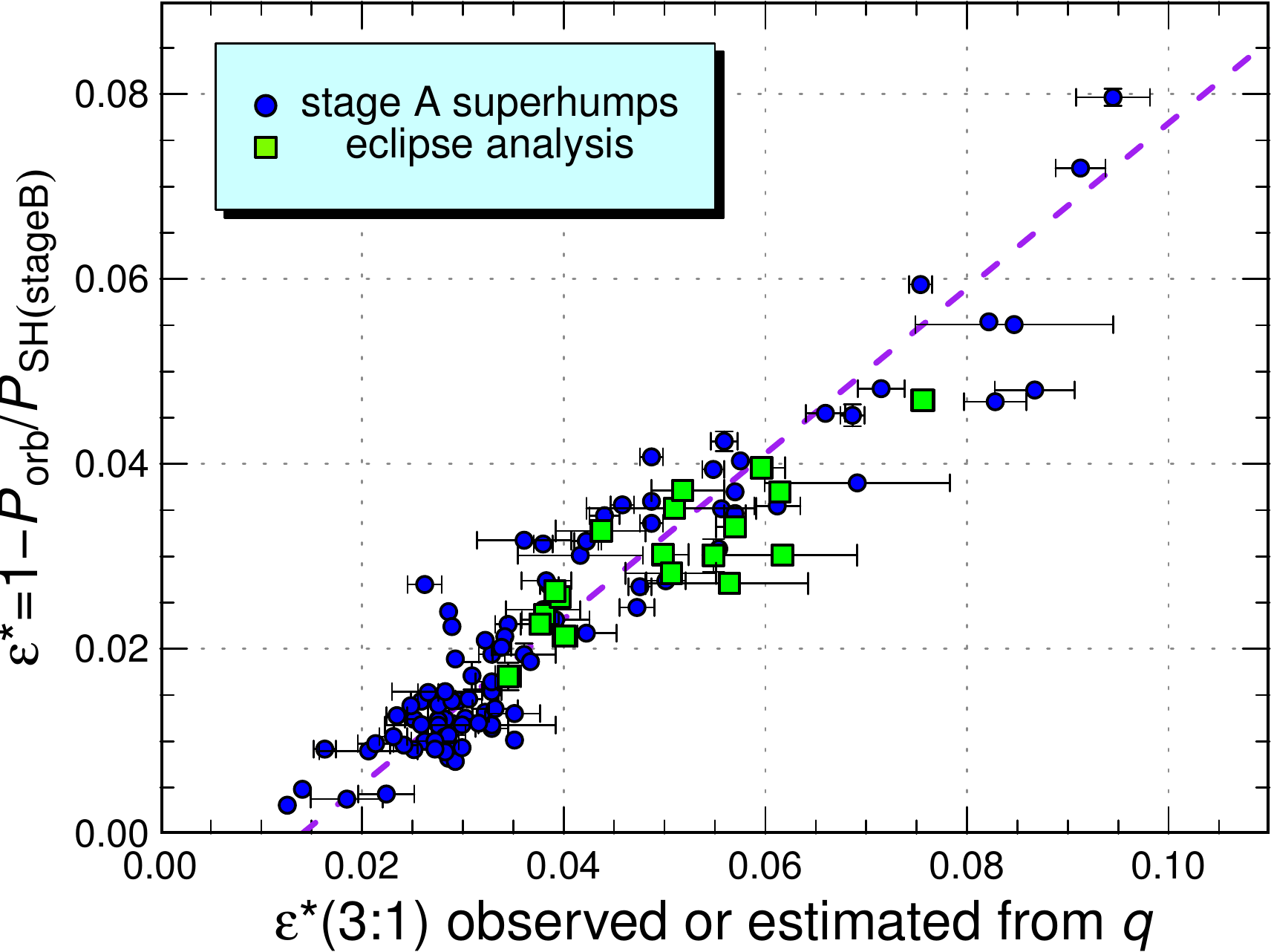}
  \end{center}
  \caption{Relation between $\epsilon^*$ of stage B superhumps and
     dynamical precession rate = $\epsilon^*$(3:1) observed or
     estimated from $q$.
     The dashed line indicates a linear fit to the data.}
  \label{fig:p31epsb2new}
\end{figure*}

\begin{figure*}
  \begin{center}
    \includegraphics[width=16cm]{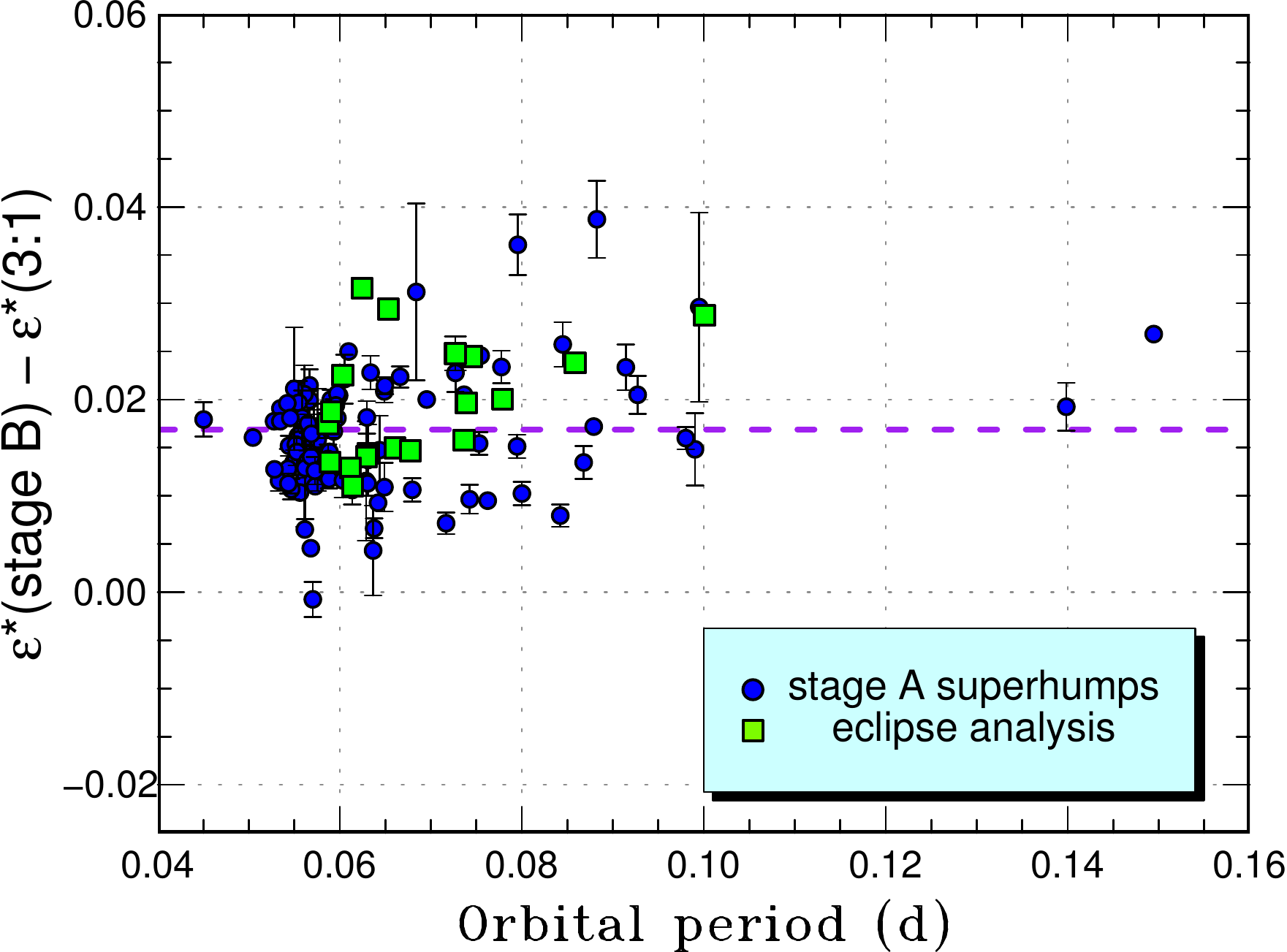}
  \end{center}
  \caption{Relation between $P_{\rm orb}$ and pressure effect
     working on stage B superhumps.
     The dashed line indicates the relation equation
     (\ref{equ:p31epsb2newconst}).}
  \label{fig:p31epsbdiff}
\end{figure*}

\begin{figure*}
  \begin{center}
    \includegraphics[width=16cm]{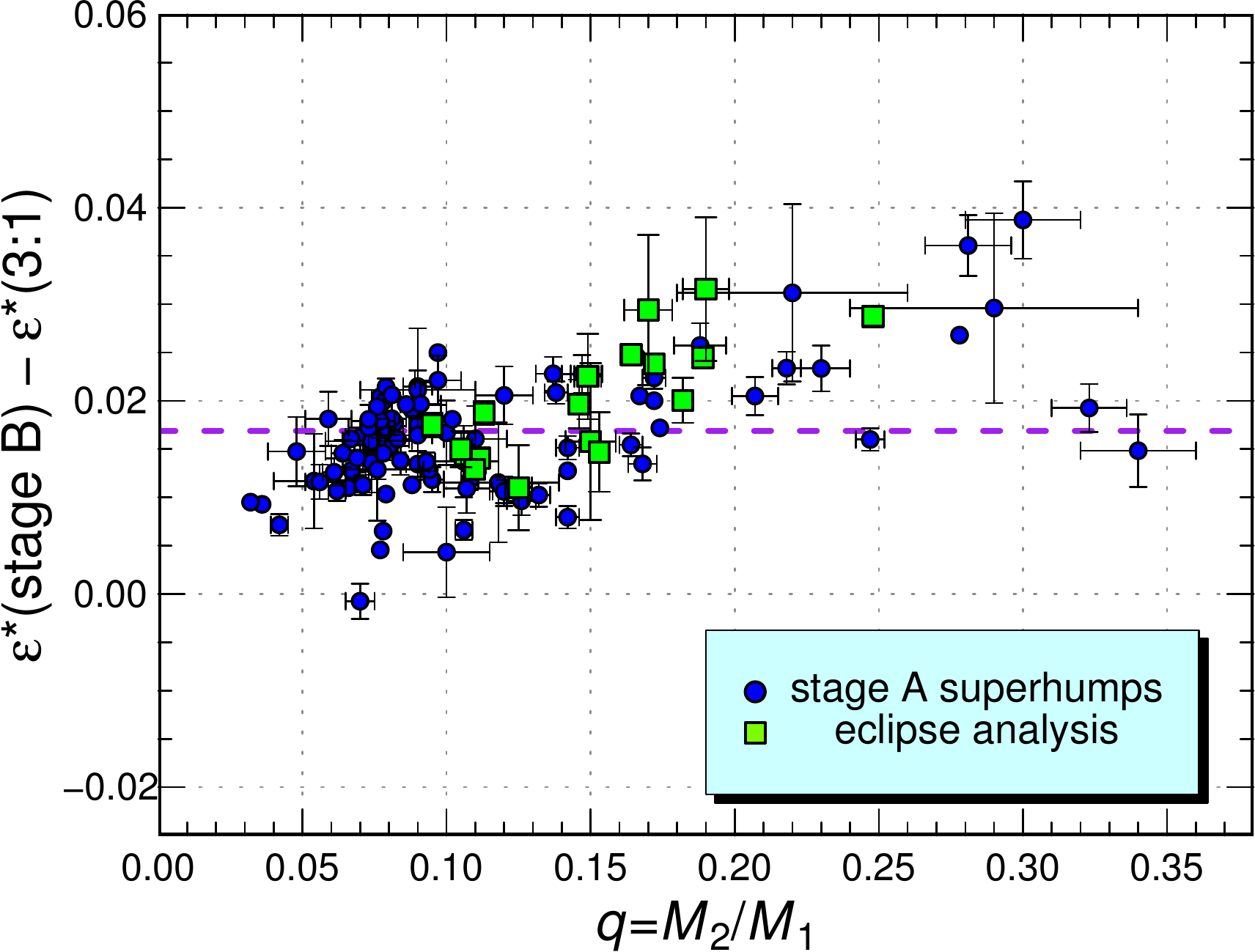}
  \end{center}
  \caption{Relation between $q$ and pressure effect
     working on stage B superhumps.
     The dashed line indicates the relation equation
     (\ref{equ:p31epsb2newconst}).}
  \label{fig:p31epsbdiffq}
\end{figure*}

   The updated relation is shown in figure \ref{fig:p31epsb2new}
(updated version of figure 9 in \cite{kat13qfromstageA};
$\epsilon^*$ is used instead of $\epsilon$ for stage B superhumps).
The linear relation is now very clear (note that the measurements
of stage A and stage B superhumps are independent).
The updated relation is
\begin{equation}
\label{equ:p31epsb2new}
\epsilon^*\math31=0.0140(11)+1.11(4)\epsilon^*\stageBmath.
\end{equation}
This relation can be safely used in all SU UMa-type dwarf novae.
For user's convenience I provide table \ref{tab:epsb}
both in reference to $\epsilon$ and $\epsilon^*$.
\textbf{However, when stage A superhumps are
well observed, do net rely on stage B and
use table \ref{tab:stageaepsq}}.
The coefficient for $\epsilon^*\stageBmath$ in equation
(\ref{equ:p31epsb2new}) is close to unity, and if I assume
it to be 1, the equation becomes
\begin{equation}
\label{equ:p31epsb2newconst}
\epsilon^*\math31=0.0169(6)+\epsilon^*\stageBmath.
\end{equation}
This equation is based on the assumption that
$\omega_{\rm pressure}$ for stage B superhumps is constant
regardless of $q$ or $\epsilon$.  The small deviation
of the coefficient from unity in equation (\ref{equ:p31epsb2new})
suggests that this is not a bad assumption.
Two figures (\ref{fig:p31epsbdiff}, \ref{fig:p31epsbdiffq})
show this relation.  The result looks very different from
figure 1 in \citet{sma20SHprecession}.
This was because \citet{sma20SHprecession} used different
classes objects including (nearly) steady state AM CVn stars
and novalike stars, and objects with large errors.
It is now apparent that the mass ratio of AM CVn
\citep{roe06amcvn} used in \citet{sma20SHprecession}
had a large uncertainty.  This is one of the reasons why
I basically did not use the $q$ values from Doppler tomography
for comparing the stage A superhump method and
the eclipse modeling method
(sections \ref{sec:data}, \ref{sec:directcomp}).

   Equation (\ref{equ:p31epsb2new}) would be meaningful
only when stage A superhump are not well observed.
Although this formula depends on an experimental calibration,
it is expected to be still better than Patterson's formulae
in that it deals with the pressure effect properly.
The calibration was done only for hydrogen-rich dwarf novae;
it is not known how the strength the pressure effect 
affects the relation other than in hydrogen-rich dwarf novae
(see e.g. \cite{pea07amcvnSH}).
Please note that $\epsilon^*\stageBmath$ or $\epsilon\stageBmath$
is not zero for $q$=0.  $\epsilon\stageBmath$
can be even negative (i.e. $P_{\rm SH}$ can be
shorter than $P_{\rm orb}$; these superhumps are
``negative superhumps'' by definition!?)
in systems with $q < 0.04$.

\begin{table*}
\caption{Relation between $\epsilon$ of stage B superhumps and $q$.
Note that this relation was experimentally calibrated and not
analytically derived.  $q<0$ is not realistic, but is given
to estimate the $\epsilon$ and $\epsilon^*$ values for $q=0$.}
\label{tab:epsb}
\begin{center}
\begin{tabular}{ccc|ccc|ccc|ccc}
\hline\hline
$\epsilon$ & $\epsilon^*$ & $q$ & $\epsilon$ & $\epsilon^*$ & $q$ & $\epsilon$ &$\epsilon^*$ & $q$ & $\epsilon$ & $\epsilon^*$ & $q$ \\ 
\hline
\xxinput{epsb.inc}
\hline
\end{tabular}
\end{center}
\end{table*}

\begin{figure*}
  \begin{center}
    \includegraphics[width=16cm]{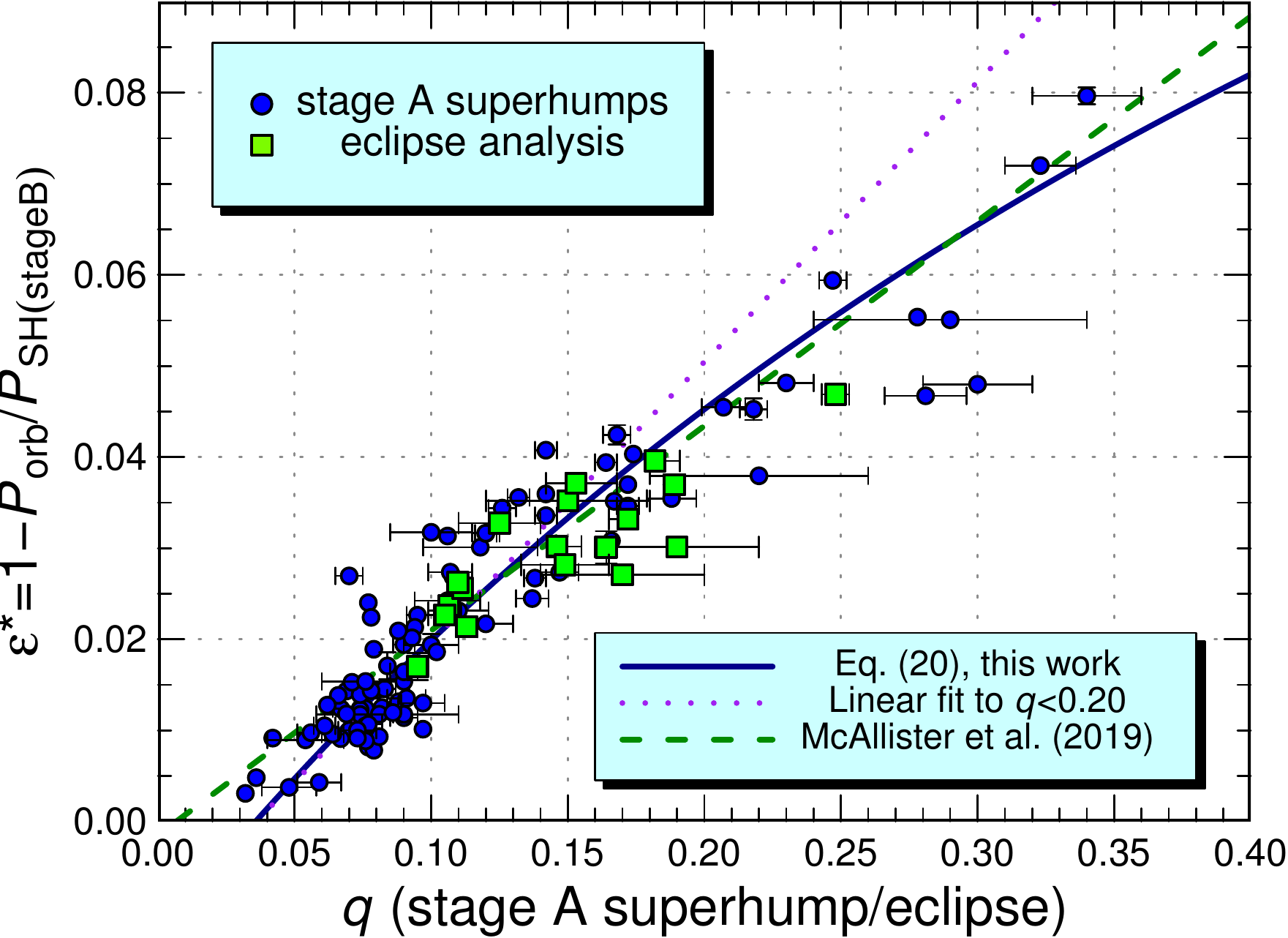}
  \end{center}
  \caption{Relation between $\epsilon^*$ of stage B superhumps and $q$.
     The dotted line indicates a linear fit to the data for $q < 0.20$
     [corresponding to linear regressions used
     in \citet{kni06CVsecondary}, but limited in the range].
     The dashed curve indicates a linear relation using $\epsilon$
     in equation (\ref{equ:mca19stageb}) = equation (2) in
     \citet{mca19DNeclipse}.
     The solid curve represents the relation in
     equation (\ref{equ:p31epsb2new}).}
  \label{fig:qepsb2new}
\end{figure*}

\begin{figure*}
  \begin{center}
    \includegraphics[width=16cm]{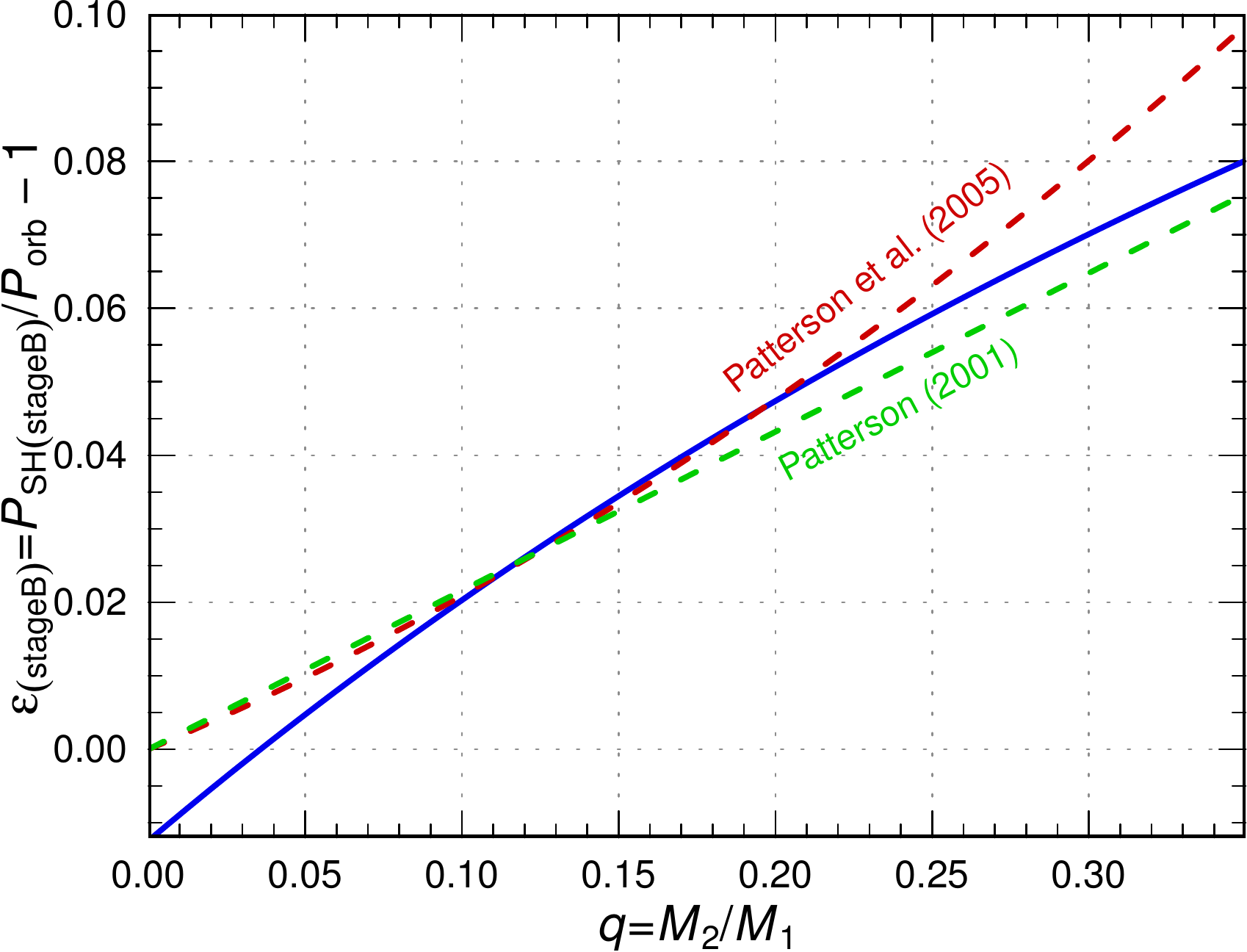}
  \end{center}
  \caption{Estimated relation between $\epsilon$ of stage B superhumps
     and $q$. The blue curve (note that it has a curvature)
     represents equation (\ref{equ:p31epsb2new})
     on the $\epsilon$-$q$ plane.
     Patterson's formulae systematically give smaller $q$ values
     for $q < 0.1$.}
  \label{fig:qepsstageb}
\end{figure*}

   Since Knigge's group published a slightly different
form of calibration
(using $\epsilon$ instead; see section \ref{sec:evol}
for more details) between stage B superhumps
and $q$, I provide a comparison of this type of formula
using the same data set (figure \ref{fig:qepsb2new}).
There is a clear tendency of deviation in the region of
$q > 0.20$.  For systems with $q < 0.20$, the relation
(dashed line in figure \ref{fig:qepsb2new}) is
\begin{equation}
\label{equ:stagebtoq}
q=0.036(4)+3.25(17)\epsilon^*\stageBmath.
\end{equation}
Note that the calibration is valid only in the range of
$0.003 < \epsilon\stageBmath < 0.045$.
The reason for the deviation for large $q$ is evident:
the precession rate at the radius of the 3:1 resonance,
$\epsilon^*\math31$, is not a linear function of $q$
(see figure \ref{fig:qeps31}) and a regression should not be
done between $\epsilon$ (or $\epsilon^*$) and $q$.
The relation between this calibration with Patterson's formulae
is shown in figure \ref{fig:qepsstageb}.
Patterson's formulae systematically give smaller $q$ values
for $q < 0.1$ because pressure effect was not properly
considered.
The advantage of the treatment in equation (\ref{equ:p31epsb2new})
or figure \ref{fig:p31epsb2new} over Knigge-type treatment
is now also obvious.

\subsection{Application to other classes of binaries}

   There have been applications of the stage A superhump method
to AM CVn stars \citep{kat14j0902,iso16crboo,iso19nsv1440,han21amcvn},
but they are not treated in this paper.
There has been the first application of the stage A superhump method
to the classical black-hole X-ray transient V3721 Oph = ASASSN-18ey =
MAXI J1820$+$070 \citep{nii21asassn18ey}.
This stage A superhump method is expected to be widely used in
black-hole X-ray transients in future in determining
the black hole masses (this is complimentary to radial-velocity studies
combined with ellipsoidal modeling in quiescence),
since the stage A superhump method is not affected
by inclinations, which usually have large uncertainties.

\section{The Data for Comparison of the Eclipse Modeling Method
and Stage A Superhump Method}\label{sec:data}

   The $q$ values determined by the stage A superhump method
are listed in table \ref{tab:stageaq}.  V627 Peg \citep{Pdot7}
is not included in this list since $P_{\rm orb}$ was not
apparently determined reliably.
The $q$ values determined by the modern eclipse method
are listed in table \ref{tab:eclq}.  Only hydrogen-rich
dwarf novae are treated in this table.

   By ``modern eclipse method'', I mean detailed modeling
of quiescent eclipses using modern equipment such as
\textsc{Ultracam} \citep{dhi01ULTRACAM,dhi07ULTRACAM}.
Such a model was described as a form of decomposition
of the eclipse light curve in \citet{woo86zcha}.
Modern treatments can be found such as
in \citet{fel04xzeridvuma,sav11CVeclmass}.
The model \textsc{lcurve}, which was a generalization of
the code used in \citet{hor94oycarHST},
was described in \citet{cop10ippeg}
(see also \cite{pzr09SDSSpostCEbinary}).
\citet{fel04xzeridvuma} used
the classical least $\chi^2$ finding algorithm
\textsc{Amoeba} (downhill simplex, \cite{NumericalRecipesFortran1986})
in determining the parameters.
In the later series of analysis by the same group, such as
\citet{sou09j1006,cop10ippeg,sav11CVeclmass},
this was taken over by the modern tool
Markov-Chain Monte Carlo (MCMC) method \citep{for06MCMC,gre07MCMC}
and the results by the MCMC method are probably more
reliable \citep{sav11CVeclmass}.  \citet{mca17asassn14ag}
employed a parallel-tempered MCMC sampler
\citep{ear05ptMCMC,for13emcee}.
This detailed modeling of quiescent eclipses
is now considered as the golden standard and $q$ values
derived by old methods or by radial-velocity studies,
which have large uncertainties, are excluded from
the analysis in this paper.  I included HT Cas
\citep{hor91htcas} despite that the $q$ value was
measured by the traditional method since there is
no refined modern measurement (the hot spot is not
always present in HT Cas; see \cite{fel05gycncircomhtcas}).
I also included EX Dra \citep{fie97exdra} which had
a relatively well-resolved eclipse light curve.
The $q$ value of 0.59(17) for SDSS J075059.97$+$141150.1
\citep{sou10SDSSeclCV} is too large for
$P_{\rm orb}$=0.09317~d and is not used.

   I apologize if there are omissions, although I extensively
used NASA's Astrophysics Data System (ADS).
I used only already published results in these tables.

\begin{center}
\begin{longtable}{lcccc}
\caption{Mass ratios determined by stage A superhump method.}\label{tab:stageaq} \\
\hline\hline
Object & $P_{\rm orb}$ (d) & $q$ & Error & References \\
\hline
\endfirsthead
\caption{Mass ratios determined by stage A superhump method (continued).} \\
\hline\hline
Object & $P_{\rm orb}$ (d) & $q$ & Error & References \\
\hline
\endhead
\hline
\endfoot
\endlastfoot
\xxinput{totab.inc}
\hline
\end{longtable}
\end{center}

\begin{center}
\begin{longtable}{lcccc}
\caption{Mass ratios determined by modern eclipse modeling.}\label{tab:eclq} \\
\hline\hline
Object & $P_{\rm orb}$ (d) & $q$ & Error & References \\
\hline
\endfirsthead
\caption{Mass ratios determined by modern eclipse modeling (continued).} \\
\hline\hline
Object & $P_{\rm orb}$ (d) & $q$ & Error & References \\
\hline
\endhead
\hline
\endfoot
\endlastfoot
\xxinput{knownq.inc}
\hline
\end{longtable}
\end{center}

\section{Comparison between Eclipse Modeling Method and Stage A Superhump Method}\label{sec:directcomp}

   Only three objects (HT Cas, Z Cha and IY UMa) are common to
tables \ref{tab:stageaq} and \ref{tab:eclq}.
XZ Eri, V1239 Her and DV UMa, which were used 
in \citet{kat13qfromstageA},
have been excluded from this list since it has become
apparent that stage A superhumps were observed only
for a short segment [see figure 87 in \citet{Pdot} and
figures 52 and 153 in \citet{Pdot4};
the values given in \citet{Pdot,Pdot4} should be regarded
as the lower limit].
The adopted objects have $P_{\rm orb}$ longer than 0.07~d
and they are not very favorable targets for
the stage A superhump method, except HT Cas, which was
very extensively observed during the 2017 superoutburst
(see figure 9 in \cite{Pdot3}).
This is because the accuracy of $q$ by the stage A superhump method
depends on the duration of stage A, which is short in
long-$P_{\rm orb}$ systems.
The presence of eclipses also complicates determination
of the superhump period (especially when eclipses overlap
superhump maxima, the superhump period cannot be reliably
determined).  Due to these factors,
this direct comparison between the results of
the eclipse modeling method and the stage A superhump method
is not very fruitful as of now.
This situation is expected to be improved by an accumulation
of continuous short-cadence observations such as
by the Transiting Exoplanet Survey Satellite (TESS).

   The lack of short-period objects common to these
two methods was due to the low frequency of
outbursts in the objects analyzed by
the eclipse modeling method.
This is because eclipse modeling is suitable for
low-mass transfer systems (such as WZ Sge stars)
which is less disturbed by flickering and by the light
from the accretion disk (and probably because hunting
brown-dwarf secondaries was one of the primary targets
by the eclipse modeling method).
The stage A superhump method requires
observations during superoutbursts.  Such outbursts
in WZ Sge stars usually occur once in decades
and time will tell whether the stage A superhump method
gives the same $q$ values for individual WZ Sge stars
measured by the eclipse modeling method.

\begin{figure*}
  \begin{center}
    \includegraphics[width=14cm]{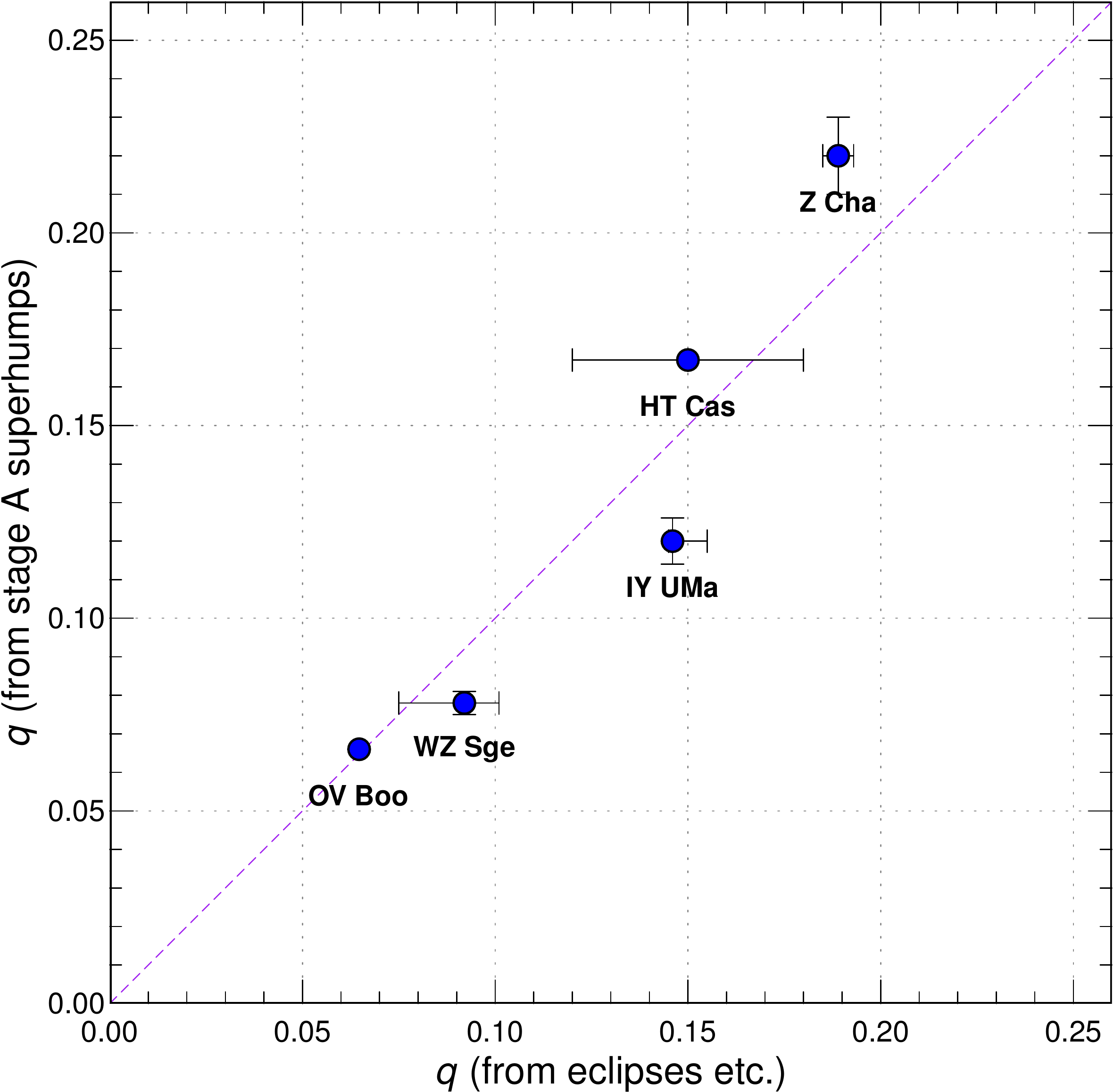}
  \end{center}
  \caption{Comparison of $q$ measured by the eclipse modeling method
  (abscissa) and $q$ estimated from the stage A superhump method
  (ordinate).  The $q$ value for WZ Sge was determined spectroscopically
  by \citet{ste07wzsge}.  We also used unpublished data for OV Boo
  \citep{ohn19ovboo}.}
  \label{fig:stageacomp}
\end{figure*}

   I nevertheless provide table \ref{tab:stageawithq} and
figure \ref{fig:stageacomp} for comparison.  In order to
increase the sample size, I used the dynamical constraints
for WZ Sge by spectroscopic observations by \citet{ste07wzsge}.
\citet{ste07wzsge} implied $q$=0.092(8) based on
the gravitational redshift, which would give a smaller
error bar shown in this figure.  \citet{ste07wzsge}
also stated that $q$=0.050(15) from superhump observations
by \citet{pat05SH} is too small and that this cannot
be a reliable independent determination.  This is exactly
what I described in section \ref{sec:stageAmethod}
(Patterson's formulae give systematically small $q$
values for low-$q$ systems).
I also included still unpublished superhump observations
of OV Boo \citep{ohn19ovboo}, which recorded the entire
superoutburst with ideal coverage.  Let's hope that
this OV Boo paper will come into the world soon.

\section{Distributions on the $P_{\rm orb}$-$q$ Plane}

\begin{figure*}
  \begin{center}
    \includegraphics[width=16cm]{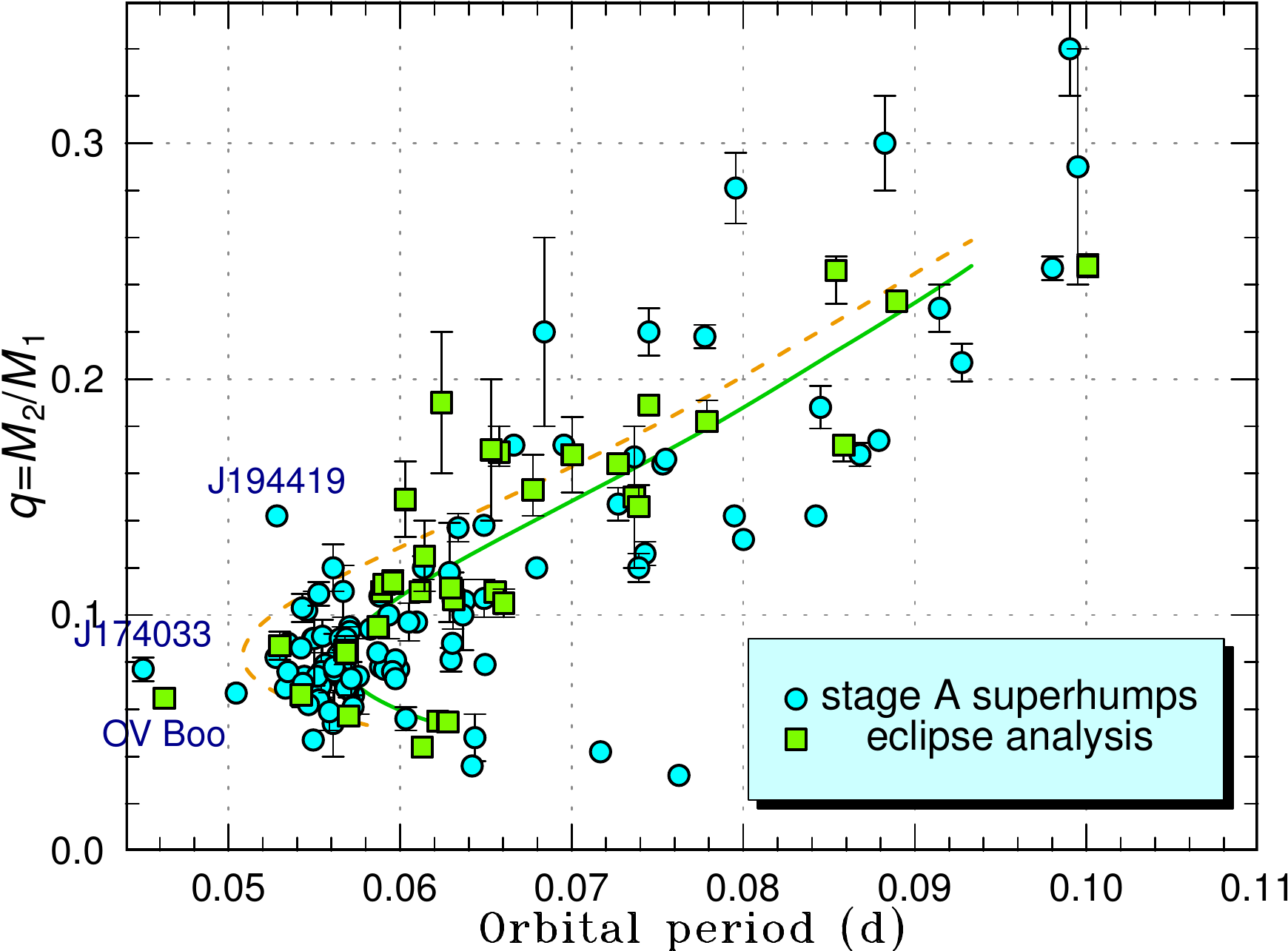}
  \end{center}
  \caption{Mass ratios ($q$) versus orbital periods ($P_{\rm orb}$)
  determined by the eclipse modeling method and
  the stage A superhump method.
  The dashed and solid curves represent the standard and optimal
  evolutionary tracks in \citet{kni11CVdonor}, respectively.
  Three unusual objects are also plotted:
  OV Boo \citep{pat17ovboo}, which is a population II CV,
  ASASSN-15po \citep{nam17asassn15po} and
  GALEX J194419.33$+$491257.0 (J194419) \citep{kat14j1944}.
  }
  \label{fig:qallnew}
\end{figure*}

   In figures \ref{fig:qalllarge} and \ref{fig:qallnew},
I show the distribution of the measurements on
the $P_{\rm orb}$-$q$ plane.
The stage A superhump method and the eclipse modeling method
give the same distribution.
There are relatively large scatters for $P_{\rm orb}$
longer than 0.07~d.  This is because long-$P_{\rm orb}$
systems are not ideal targets both for
the stage A superhump method and the eclipse modeling method.
The reason for the stage A superhump method was already
described in section \ref{sec:directcomp}.
The duration of stage A is short for these systems
and stage A tends to be contaminated by stage B
(giving smaller $q$ values), if observations are not
made sufficiently early (this problem can be avoided
if continuous short-cadence observations such as by TESS
become available).
The reason for the eclipse modeling method is that these
objects have high mass-transfer rates and the profiles
of eclipses tend to be blurred or the light curve
is dominated by flickering, although this is alleviated
by using Gaussian process modeling \citep{mca17asassn14ag}.
This eclipse modeling also requires the presence
of the hot spot.  In \citet{fel05gycncircomhtcas},
$q$ values were not determined for IR Com and HT Cas
due to the lack of the hot spot.

   The two unusual objects plotted on figure \ref{fig:qalllarge}
are ASASSN-15po and GALEX J194419.33$+$491257.0.
The former is an object having $P_{\rm orb}$
below the period minimum and may not be on
the standard evolutionary path of CVs
\citep{nam17asassn15po}.  The latter is an unusually
active SU UMa star with very short recurrence times
and is suspected to be a CV with a stripped core
evolved secondary \citep{kat14j1944}.
Additional two unusual object plotted on figure \ref{fig:qallnew}
are CRTS J174033.4$+$414756, which is also considered
to be a CV with a stripped core
evolved secondary \citep{cho15j1740,ima18j1740}
and OV Boo, which is considered to be a population II CV
\citep{pat17ovboo}.

\begin{table*}
\caption{Comparison of $q$ values determined by the eclipse modeling method
and the stage A superhump method.}\label{tab:stageawithq}
\begin{center}
\begin{tabular}{cccccc}
\hline\hline
Object & $P_{\rm orb}$ (d) & $q$ (eclipse) & Error & $q$ (stage A) & Error \\\hline
OV Boo & 0.046258 & 0.0647 & 0.0018 & 0.066 & 0.001 \\
WZ Sge & 0.056688 & 0.092 & $+$0.009/$-$0.017 & 0.078 & 0.003 \\
HT Cas & 0.073647 & 0.15 & 0.03 & 0.167 & 0.002 \\
IY UMa & 0.073909 & 0.146 & $+$0.009/$-$0.001 & 0.120 & 0.006 \\
Z Cha  & 0.074499 & 0.189 & 0.004 & 0.22 & 0.01 \\
\hline
\end{tabular}
\end{center}
\end{table*}

\begin{flushleft}
\large{\textbf{Conclusion of this section:}}
\end{flushleft}

   Both figures now clearly show that \textbf{the eclipse modeling
method and the stage A superhump method give the same distribution}.
This means that \textbf{the stage A superhump method is as reliable
as the eclipse modeling method}.

\section{CV Evolution and Period Minimum}\label{sec:evol}

   It is well known that the distribution of CV has a period minimum
\citep{pac81CVGWR}.  Around this point during the course of
the CV evolution, the $P_{\rm orb}$ increases mainly due to
two reasons: the thermal time-scale (Kelvin-Helmholtz time)
of the secondary exceeds the mass-transfer time-scale and
the mass-radius relation is reversed for degenerate dwarfs
[modern works have shown that the picture is a bit more complex,
see e.g. \citet{kol99CVperiodminimum,ara05MCV,kni11CVdonor}].
For hydrogen-rich CVs, \citet{pac81CVGWR} suggested a period
of 74~min (0.051~d) for $M_1$=0.5$M_\odot$ and 87~min (0.060~d)
$M_1$=1.0$M_\odot$.  \citet{pac81CVGWR} argued that these periods
are remarkably close to the observed period of 81~min (0.056~d)
for WZ Sge.  Although early models indicated the period
between 60 and 80~min (0.042 and 0.060~d)
\citep{pac81CVGWR,rap82CVevolution,pac83CVevolution},
later refinement of the model yielded significantly shorter
values of 65--70~min (0.045--0.049~d)
\citep{kol99CVperiodminimum,how01periodgap}.
This was apparently not in agreement with the observation
(e.g. \cite{kol93CVpopulation}), and this problem was
called the ``period minimum problem''.  Population synthesis
studies expected that most CVs have already reached
the period minimum \citep{kol93CVpopulation,how97periodminimum}
and that here should be a heavy accumulation of systems around
the period minimum (period spike or period minimum spike),
since the drop in the mass-transfer rate slows
down the CV evolution.
At that time, such a spike was not apparent and this disagreement
was called the ``period spike problem''
\citep{kol99CVperiodminimum,ren02CVminimum}.
There was an idea that the mass-transfer rates rapidly decrease
as CVs approach the period minimum and that they become
increasingly difficult to find by detections of outbursts.
\citet{uem10shortPCV} assumed that the recurrence time
of outbursts near the period minimum follows an exponential law
and explored whether there could be a ``missing'' population
of CVs below the shortest known orbital period.

   Systematic spectroscopic and photometric follow-up observations
of the CV candidates selected by the Sloan Digital Sky Survey
(SDSS: \cite{SDSS})
(\cite{szk02SDSSCVs,szk03MCVSDSS,szk03SDSSCV2,szk04SDSSCV3,szk05SDSSCV4,
szk06SDSSCV5,szk07SDSSCV6,szk09SDSSCV7,szk11SDSSCV8,szk18CVs2,
sou06SDSSCV,sou07SDSSCV2,sou08CVperiod,sou08j2205,
sou09j1006,sou10SDSSeclCV,sou10j0039,sou15SDSSCV9,wol03j1327,hom06SDSSCV,
dil08SDSSCV,sch08SDSSMCV,hil09SDSSCVs,ski11j1544,wou04j1610,wou10CVperiod,
wou04CV4,wou12SDSSCRTSCVs,reb14j0011,tho15SDSSCVs};
this list includes other types of CVs not treated in this paper)
answered these issues.  \citet{gan09SDSSCVs}
collected SDSS-selected and other CVs and clarified
the presence of the period spike.  Selection of CVs using
the SDSS was advantageous in that it did not require outburst
detections and it was considered that this sample was less
biased than the traditional one.  There remained, however,
a possibility that systems with very low-mass transfer rates
can have colors indistinguishable from white dwarfs and their
number might have been underestimated (subsubsection
5.2.1 in \cite{gan09SDSSCVs}; \cite{kat12DNSDSS}).
\citet{gan09SDSSCVs} selected GW Lib
[$P_{\rm orb}$=0.05332(2) d or 76.78~min] as
the shortest-period ``standard'' hydrogen-rich CV.
\citet{gan09SDSSCVs} reported the period spike of
82.4(7)~min = 0.0572(5)~d with a FWHM of 5.7$\pm$1.7~min
= 0.0040(12)~d.  Considering the width of the period spike,
a period of 80~min = 0.0556~d was considered to be
an estimate for the period minimum.  \citet{gan09SDSSCVs}
stated that the theoretically calculated period minimum is
too short by about 10~min.  This difference could be
reconciled if either the orbital braking is about four
times the value provided by gravitational wave radiation
\citep{kol99CVperiodminimum}, or if the theoretical models
underestimate the stellar radius for a given mass by
about 20\% \citep{bar03CVminimumperiod}.

   \citet{kni06CVsecondary} used superhumping CVs
(SU UMa stars and permanent superhumpers) and eclipsing CVs
to explore the mass-radius relation for the secondary stars
in CVs.  \citet{kni06CVsecondary} reported a period minimum
of 76.2$\pm$1.0~min = 0.0529(7)~d.  
It might be worth noting that \citet{kni06CVsecondary}
used the originally calibrated formula between $q$
and $\epsilon$ (stage B for SU UMa stars):
\begin{equation}
q(\epsilon) = (0.114 \pm 0.005) + (3.97 \pm 0.41) \times (\epsilon - 0.025),
\end{equation}
using the same set of calibrators as in \citet{pat05SH}.
This formula is better than equation (\ref{equ:pat05eq}),
the relation by \citet{pat05SH}, in that it allows
$q \ne 0$ at $\epsilon=0$.  For example, $q$=0.015
is obtained for $\epsilon=0$ by this equation (compare with
figure \ref{fig:qepsstageb}).

   \citet{kni11CVdonor} studied the evolution of CVs
in detail considering various effects.  Using the material
in \citet{kni06CVsecondary}, \citet{kni11CVdonor} concluded
that the best-fit model for short-$P_{\rm orb}$ systems
requires angular momentum losses (AML) 2.47$\pm$0.22 larger
than what is expected by gravitational wave radiation
(optimal evolutionary track).  This optimal evolutionary track
and standard evolutionary track (angular momentum losses
exactly by gravitational wave radiation) are plotted on
figures \ref{fig:qallnew} and \ref{fig:qalllarge}.
\citet{kni11CVdonor} derived the theoretical period minimum
of 73.2~min = 0.0508~d and 81.8~min = 0.0568~d for the standard
and optimal evolutionary tracks, respectively.

   The value of the period minimum depends on the definition.
\citet{Pdot9} used $P_{\rm orb}$ values estimated from
$P_{\rm SH}$ for all SU UMa/WZ Sge stars and derived
a sharp cut-off at 0.052897(16).  The used relations between
$P_{\rm orb}$ and $P_{\rm SH}$ (stage B or C) were
in \citet{Pdot3}, but repeated here for easier reference:
\begin{equation}
\epsilon\stageBmath = 0.000346(36)/(0.043-P_{\rm orb}) + 0.0443(21)
\label{equ:p1porb}
\end{equation}
and
\begin{equation}
\epsilon\stageCmath = 0.000273(24)/(0.044-P_{\rm orb}) + 0.0381(13).
\label{equ:p2porb}
\end{equation}
These equations have smaller residuals (1-$\sigma$ error
of 0.0003~d) compared to older compared to the one
in \citet{sto84tumen}, which had a systematic error.

   \citet{mca19DNeclipse} used all SU UMa/WZ Sge stars
as in \citet{Pdot9} and eclipsing CVs and derived
the period minimum (defined as the period spike
as in \cite{gan09SDSSCVs}) of 79.6(2)~min = 0.0553(1)~d
with a FWHM of 4.0~min = 0.003~d.
\citet{mca19DNeclipse} claimed that they derived
new Knigge-type calibrations of the relationship between $\epsilon$
and $q$ [they were probably unaware of \citet{kat13qfromstageA}],
namely,
\begin{equation}
\label{equ:mca19stageb}
q(\epsilon_{\rm B}) = (0.118 \pm 0.003) + (4.45 \pm 0.28) \times (\epsilon_{\rm B} - 0.025)
\end{equation}
and
\begin{equation}
\label{equ:mca19stagec}
q(\epsilon_{\rm C}) = (0.135 \pm 0.004) + (5.0 \pm 0.7) \times (\epsilon_{\rm C} - 0.025),
\end{equation}
where $\epsilon_{\rm B}$ and $q_{\rm B}$ represent stage B superhumps
and $\epsilon_{\rm C}$ and $q_{\rm C}$ represent stage C superhumps.
Although the relation for stage B was
greatly improved compared to \citet{pat01SH} or
\citet{pat05SH}, this relation has is inferior to
equation (\ref{equ:p31epsb2new}) since the $\epsilon$-$q$
relation is not linear (see subsection \ref{sec:stagebtoq}).
\citet{mca19DNeclipse} also wrote: ``While there is good coverage
for systems with $0.1 < q < 0.2$, more calibration systems
with $q$ outside this range are required in order to further
constrain the gradient.'' --- This is because systems
with $q < 0.1$ are usually WZ Sge stars
(or borderline SU UMa/WZ Sge stars) and they usually do not show
prominent stage C superhumps; hence the equation for stage C
superhumps is not necessary for $q < 0.1$.
In any case, the departure of the period minimum
from the theoretically calculated one was also apparent.

   \citet{wil22shortPCV} added three examples and evaluated
the mass-radius relation in \citet{kni11CVdonor}.
They also used superhumps probably as in the way as in
\citet{mca19DNeclipse}, but the details and references
were not shown.
\citet{wil22shortPCV} used our superhump periods
of MASTER OT J220559.40$-$341434.9 = ASASSN-16kr \citep{Pdot9}
and obtained $q$=0.059(7) using the relation in
\citet{mca19DNeclipse}.  \citet{wil22shortPCV} suggested
that this may be preliminary evidence that the $\epsilon$-$q$
relation may overestimate $q$ for CVs at short periods.
I looked at the data of this object again and found
that the data were not adequate for making such a comparison.
There were observations by Monard on four nights and
Hambsch on four nights, the latter with a low time resolution,
and these observations covered the final part of
the superoutburst plateau (the observations started
12~d after the initial detection of the superoutburst).
As stated in \citet{Pdot9}, most superhumps were affected
by overlapping eclipses and orbital humps and the period
should be regarded as on approximate.  The unusually
[written as ``usual'' in \citet{Pdot9}, which was a typo]
large period derivative also indicated that the period was not
well determined.  This object should not be used as
a $\epsilon$-$q$ calibrator until we have better measurements
of superhumps in the early part of a superoutburst in future.
\citet{wil22shortPCV} found a relation
between $P_{\rm ex} = P_{\rm obs,orb} - P_{\rm model,orb}$
and $M_2$, which is a measure of excess AML against
the optimal evolutionary sequence in \citet{kni11CVdonor}.
Based on the complex behavior, \citet{wil22shortPCV}
wrote: ``The `optimal' tracks add an extra source of AML that takes
the form of 1.5 times the gravitational wave breaking.
By examining the period excess between the growing set of
observed CV donor radii and models, we demonstrate that this
does not properly describe the missing AML.''
Although I did not examine the matter in detail, a systematic
departure of the (linear) $\epsilon$-$q$ relations
from the non-linear (correct) one might have complicated the problem
[figure \ref{fig:qepsb2new} suggests that the relation
in \citet{mca19DNeclipse} tends to overestimate $q$ for
$0.15 < q < 0.25$ and underestimate $q$ for $q < 0.10$].

   Upon a closer look at figure \ref{fig:qalllarge}
or figure 10 in \citet{mca19DNeclipse},
it is apparent that the majority of objects on the $P_{\rm orb}$-$q$
(or $P_{\rm orb}$-$M_2$) plane near the period minimum
have shorter $P_{\rm orb}$ than the optimal evolutionary track
in \citet{kni11CVdonor}, but still longer than
the standard evolutionary track.
It is also apparent that $P_{\rm orb}$ values around
the period minimum are widespread and some of the objects
apparently have $P_{\rm orb}$ longer than
the optimal evolutionary track.  Although the presence of
objects with $P_{\rm orb}$ longer than
the optimal evolutionary track is more apparent in $q$ values
measured from superhumps, there are indeed some systems measured
by the eclipse modeling method and they look like to be real.
This suggests that the ``period minimum'' is more widespread
(in $P_{\rm orb}$) than has been thought.
This implies that a given object passes a broad range
(such as 0.052--0.060~d) of the period minimum rather than
evolving through a narrow, fixed period.  The reason of
the spread is unknown.  The errors in $q$ estimates are
unlikely the main source, since the spread in $P_{\rm orb}$
is least affected by the errors of $q$ near the period minimum.
Different chemical compositions of the secondaries would
be a cause.
Different degrees of AML other than gravitational wave radiation
may be the cause, although consequential angular momentum loss
(CAML) such as caused by nova eruptions \citep{sch16CVAML}
is expected to be less important around the period minimum
\citep{mca19DNeclipse}.

\begin{figure*}
  \begin{center}
    \includegraphics[width=16cm]{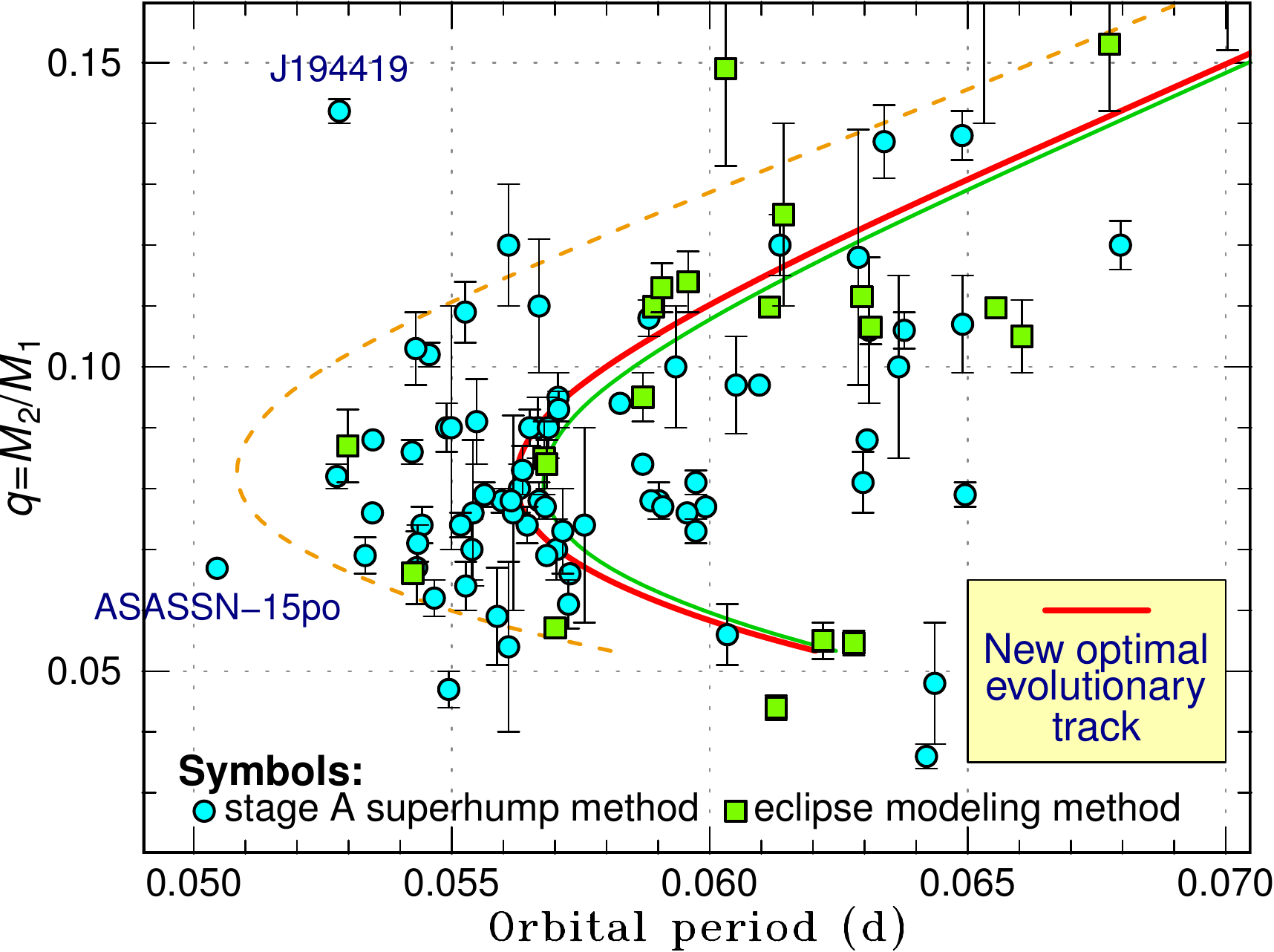}
  \end{center}
  \caption{New optimal evolutionary track (red solid curve)
  plotted on the relation between mass ratios ($q$) and
  orbital periods ($P_{\rm orb}$)
  determined by the eclipse modeling method and the stage A superhump
  method, enlargement around the period minimum.
  The other symbols are the same as in figures \ref{fig:qalllarge}
  and \ref{fig:qallnew}.
  }
  \label{fig:qalllargenew}
\end{figure*}

   In order to obtain the new ``optimal'' evolutionary track
around the period minimum, I linearly interpolated the standard
and optimal evolutionary tracks in \citet{kni11CVdonor}.
I write the two tracks with functional forms of
\begin{equation}
P_{\rm orb}{\rm(standard,optimal)} = f_{\rm standard,optimal}(q),
\end{equation}
using tables 3 and 4 in \citet{kni11CVdonor}.
I then minimized
\begin{equation}
\sum_i \Bigl[P_{{\rm orb},i}-\bigl\{c f_{\rm optimal}(q_i) + (1-c)f_{\rm standard}(q_i)\bigl\}\Bigl]^2
\end{equation}
by changing $c$, where $P_{{\rm orb},i}$ and $q_i$ represents values
for individual objects.
$c=0$ corresponds to the standard evolutionary track and
$c=1$ corresponds to the optimal evolutionary track
in \citet{kni11CVdonor}.
I have excluded ASASSN-15po, CRTS J174033.4$+$414756 and
GALEX J194419.33$+$491257.0 (for the stage A superhummp
method) and OV Boo (for the eclipse modeling method)
and the samples were limited
to $q>0.05334$ (lower limit of the evolutionary track),
$q<0.16$ and $P_{\rm orb}<0.07\;(\rm d)$.
The results are shown in table \ref{tab:newoptimal}.
The optimal value of $c$ is in excellent agreement
between the stage A superhummp method and
the eclipse modeling method (again confirming
the reliability of the stage A superhummp method).
I adopted $c$=0.91 using the combined data.
The means that the ``new'' optimal evolutionary track
is closer to the original optimal evolutionary track
than to the standard evolutionary track.
Considering that $P_{\rm orb}$ at the period minimum
is roughly proportional to AML$^{1/3}$
[equation (16) in \citet{pac81CVevolution} or
equation (2) in \citet{pac83CVevolution}],
the angular momentum loss is estimated to be 1.9 times
larger than what is expected by gravitational wave radiation.
The period minimum on this track is 0.0562~d = 81.0~min.
The can be compared to 81.8~min = 0.0568~d for
the optimal evolutionary track by \citet{kni11CVdonor}.
The new optimal evolutionary track is shown as a red curve
in figure \ref{fig:qalllargenew}.

\begin{table*}
\caption{Parameters for new optimal evolutionary track.}\label{tab:newoptimal}
\begin{center}
\begin{tabular}{lccc}
\hline\hline
Method             & Number of objects & Optimal $c$ \\
\hline
Stage A superhumps & 70 & 0.91 \\
Eclipse modeling   & 19 & 0.91 \\
Combined           & 89 & 0.91 \\
\hline
\end{tabular}
\end{center}
\end{table*}

\section{Period Bouncers}

   There have been mainly two sources of (candidate)
period bouncers.  One is by eclipse modeling starting from
the discovery of SDSS J103533.02$+$055158.3 \citep{lit06j1035}.
The other is by superhump methods calibrated by various
authors.  \citet{pat98evolution} was probably the first
to provide a list of candidate period bouncers.
There have been a number of candidate since then and
\citet{pat01SH} provided a list of candidates from
various aspects (including those discovered by
the eclipse modeling method).
Since the establishment of the stage A superhump method
in 2013 and the discovery of double-superoutburst object
with $q$=0.04 \citep{kat13qfromstageA},
there have been a growing number of candidate
period bouncers by this method.
It became evident that the type of WZ Sge-type rebrightening
is related to the evolutionary stage (see the next
section for more details), and outburst parameters
became useful in selecting period bouncers.
A list of candidate period bouncers from superhump
observations and outburst characteristics was given
in \citet{kim16asassn15jd} and updated
in \citet{kim18asassn16dt16hg}.  I list these two references
since researchers using the eclipse modeling method
did not appear to be aware of these works, but was
properly referenced by \citet{tho20shortPCV}.

   \citet{pat01SH} estimated that the period bounce occurs
at a mass of 0.058(8)$M_\odot$, which was likely
an underestimate considering the systematic error
in the $\epsilon$-$q$ relation (section \ref{sec:stageAmethod}).
\citet{mca19DNeclipse} estimated that 30\% of their sample
(including superhumpers) appear to be post-period minimum systems.
They derived the mass of the secondary at the period bounce
to be $M_{\rm bounce}$=0.063$_{-0.002}^{+0.005}M_\odot$.
This value can be compared to the lower limit of hydrogen
burning (Kumar's limit: \cite{kum63BDmass}) of
0.076(5)$M_\odot$ \citep{hen99BDmass} (from observations)
and slightly below 0.072$M_\odot$ \citep{cha97lowmassstar}
(from theoretical calculation) for isolated
population I objects.
Using their average mass (0.81$\pm$0.02$M_\odot$)
of white dwarfs below the period gap, this corresponds to
$q$=0.078 (with an error of an order of 0.003).
Using this value and ignoring all errors, 33 out of
103 objects determined by the stage A superhump
method are expected to be post-period minimum systems.
The fraction is in very good agreement
with \citet{mca19DNeclipse}.

\begin{figure*}
  \begin{center}
    \includegraphics[width=16cm]{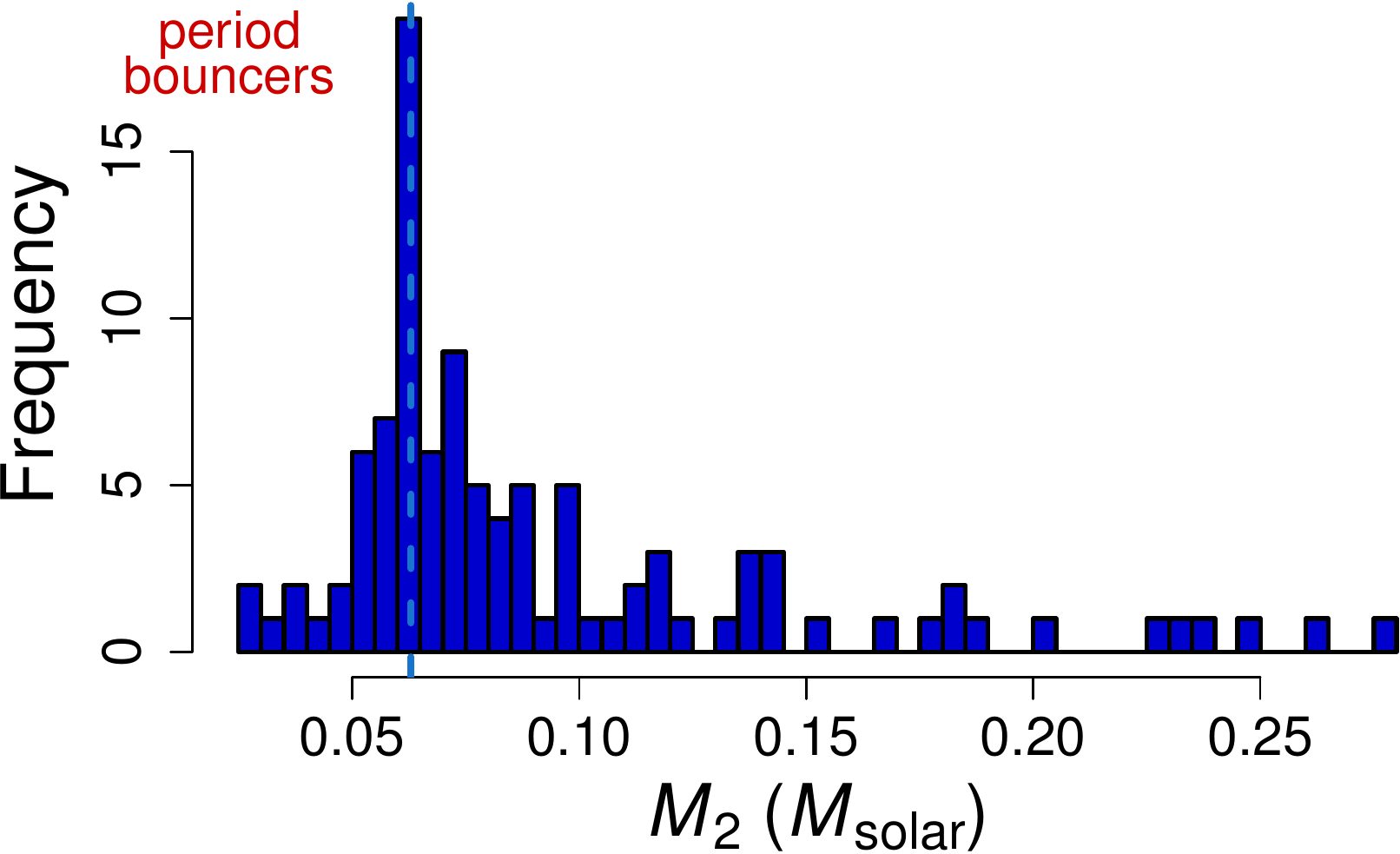}
  \end{center}
  \caption{Distribution of $M_2$ determined by the stage B
    superhump method.  An average white dwarf mass of
    0.82$M_\odot$ was assumed.  A vertical dashed line represents
    the location of the period bounce ($M_2$ = 0.063$M_\odot$).}
  \label{fig:m2hist}
\end{figure*}

   The distribution of $M_2$ determined by the stage B
superhump method, ignoring all errors, is shown in
figure \ref{fig:m2hist}.  It is worth noting the existence
of a sharp peak around the period minimum (and perhaps
near the hydrogen-burning limit).  The number density
is expected to be higher below this limit since
the evolutionary time scale becomes significantly longer.
The sharp drop of the number density below the period minimum
may be a reflection of the sharply decreasing mass-transfer
rate, which makes outbursts infrequent and difficult to
find by outburst detections.  Consequently, it would be
natural to think that there are many ``dormant''
period bouncers which still await our detections.

   The chronology of period bouncers and brown-dwarf secondaries,
together with the development of the techniques, is listed
in table \ref{tab:bouncer}.  This table would be useful
for searching which period bouncers were reported first
and what method was used for them.  This table includes
polars with brown-dwarf secondaries.  SED stands for
spectral energy distribution in the table.

\xxinput{bouncer.inc}

\section{$P_{\rm orb}$, $q$ and WZ Sge-type Rebrightenings}

\begin{figure*}
  \begin{center}
    \includegraphics[width=16cm]{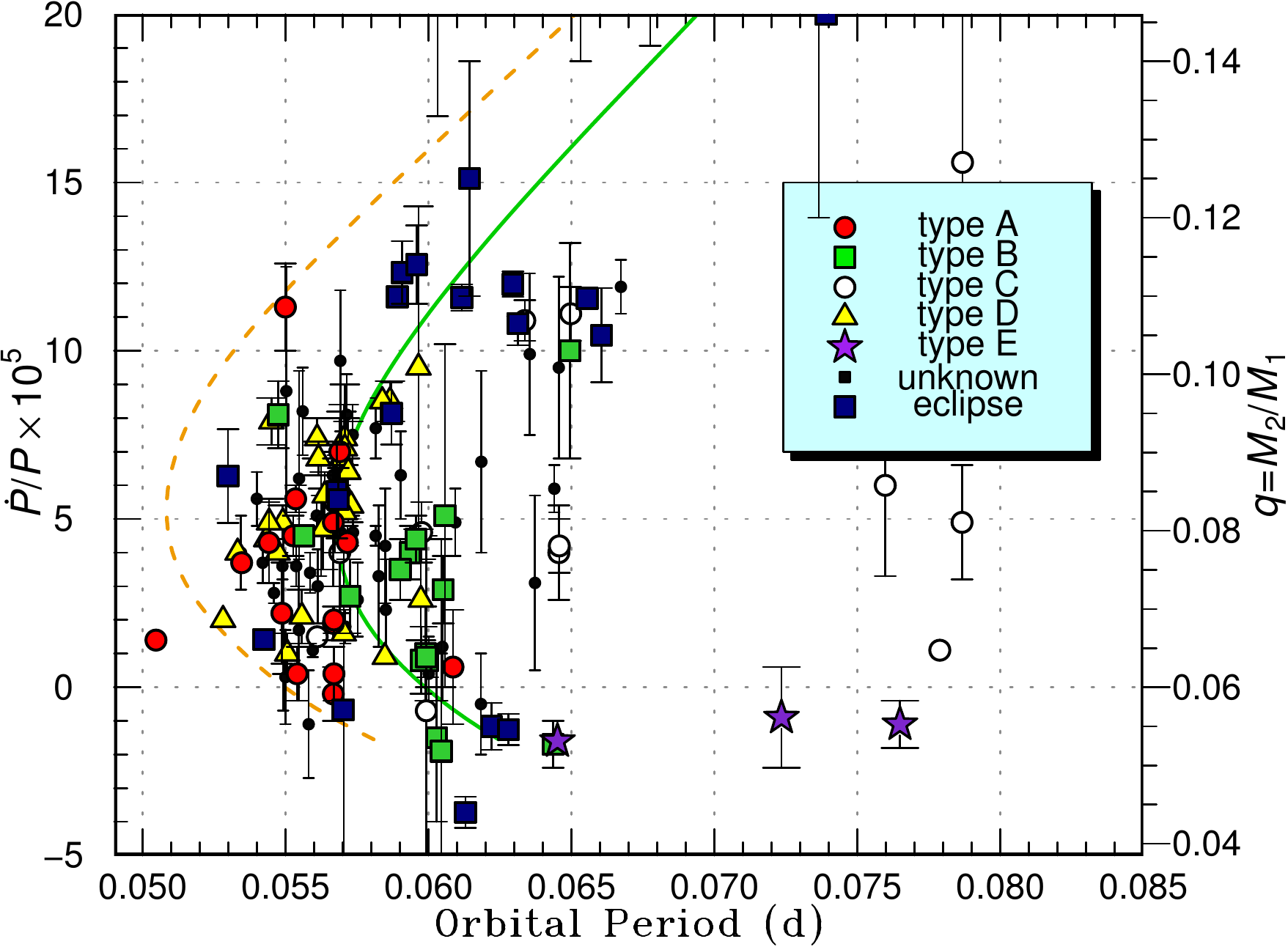}
  \end{center}
  \caption{$P_{\rm dot}$ versus $P_{\rm orb}$ for WZ Sge-type
  dwarf novae.  Symbols represent the type of outburst:
  type-A (long rebrightening, filled circles),
  type-B (multiple rebrightenings, filled squares),
  type-C (single rebrightening, filled triangles),
  type-D (no rebrightening, open circles)
  and type-E (double superoutburst, filled stars).
  On the right side, I show mass ratios estimated
  using equation (\ref{equ:pdottoq}).
  We can regard this figure as to represent
  an evolutionary diagram.
  The objects with $q$ values obtained by the eclipse modeling
  method are also plotted.
  This figure is refinement of figure 61 in \citet{Pdot9}.
  }
  \label{fig:wzpdottype}
\end{figure*}

   Most of short-$P_{\rm orb}$ and low-$q$ dwarf novae are
WZ Sge stars.  WZ Sge stars often show post-superoutburst
rebrightenings.  The rebrightenings have a variety
of morphology and they are classified into
type-A (long rebrightening), type-B (multiple rebrightenings),
type-C (single rebrightening), type-D (no rebrightening) and
type-E (double superoutburst) \citep{kat15wzsge}.
In WZ Sge stars, there is an empirical relation between
the period derivative of stage B superhumps
($P_{\rm dot} = \dot{P}/P$) and $q$
(figure 21 in \cite{kat15wzsge}).
The empirical relation [equation (6) in \cite{kat15wzsge}] is
\begin{equation}
\label{equ:pdottoq}
q = 0.0043(9)P_{\rm dot}\times 10^5 + 0.060(5).
\end{equation}
Although this relation is experimental and is calibrated
only in the range $0.04 < q < 0.12$, it is regarded as
reliable since a $P_{\rm orb}$-$P_{\rm dot}$ diagram
very clearly depicts the expected evolutionary track
(see figure 17 in \cite{kat15wzsge}).
Systems with known $P_{\rm dot}$ can be plotted on
this diagram with different marks for different types
of rebrightenings.  It has become evident that the types
of rebrightenings indicate the following evolutionary sequence:
type C $\rightarrow$ D $\rightarrow$ 
A $\rightarrow$ B $\rightarrow$ E \citep{kat15wzsge}.
The same type of figure is shown in
figure \ref{fig:wzpdottype}.\footnote{
   During the preparation of this figure, I noticed that
   the corresponding figure 61 in \citet{Pdot9} has incorrect
   labels for $q$.  The $q$ labels are corrected
   in this figure.
}
Objects with $q$ values obtained by the eclipse modeling method
are also plotted on this figure using equation (\ref{equ:pdottoq}).
It is apparent that these objects are on the same evolutionary
sequence depicted by $P_{\rm dot}$.

   The most interesting test from this figure
is whether the lowest-$q$ systems determined by
the eclipse modeling method show rebrightenings
expected by the stage A superhump method (or by
the $P_{\rm dot}$-$q$ relation).
These objects below $q$=0.06 are
ASASSN-16kr = MASTER OT J220559.40$-$341434.9,
SDSS J105754.25$+$275947.5, CRTS J052209.7$-$350530
and SDSS J103533.02$+$055158.3.
They are expected to show either type-B (multiple rebrightenings)
or type-E (double superoutburst) superoutbursts.
A superoutburst was observed only one of these four
(ASASSN-16kr, but the observational coverage was insufficient).
Observations reported to VSNET did not cover
the post-superoutburst phase.  The All-Sky Automated Survey
for Supernovae (ASAS-SN: \cite{ASASSN,koc17ASASSNLC})
data\footnote{
  $<$https://asas-sn.osu.edu/sky-patrol/coordinate/4822659d-5613-4692-beb0-5724f2401d9b$>$.
} showed at least one post-superoutburst rebrightening
on 2016 October 7 at $V$=15.5.  There were gaps of
observations and I could not confirm whether there were
more rebrightening(s).  Future observations of these
lowest-$q$ systems during superoutbursts (stage A superhumps,
of course, to obtain independent $q$ estimates)
and post-superoutburst states will be a very
interesting topic to test the evolutionary picture derived
from superhump observations.

\section{Do Some CVs Have Very Light White Dwarfs?}

\begin{figure*}
  \begin{center}
    \includegraphics[width=16cm]{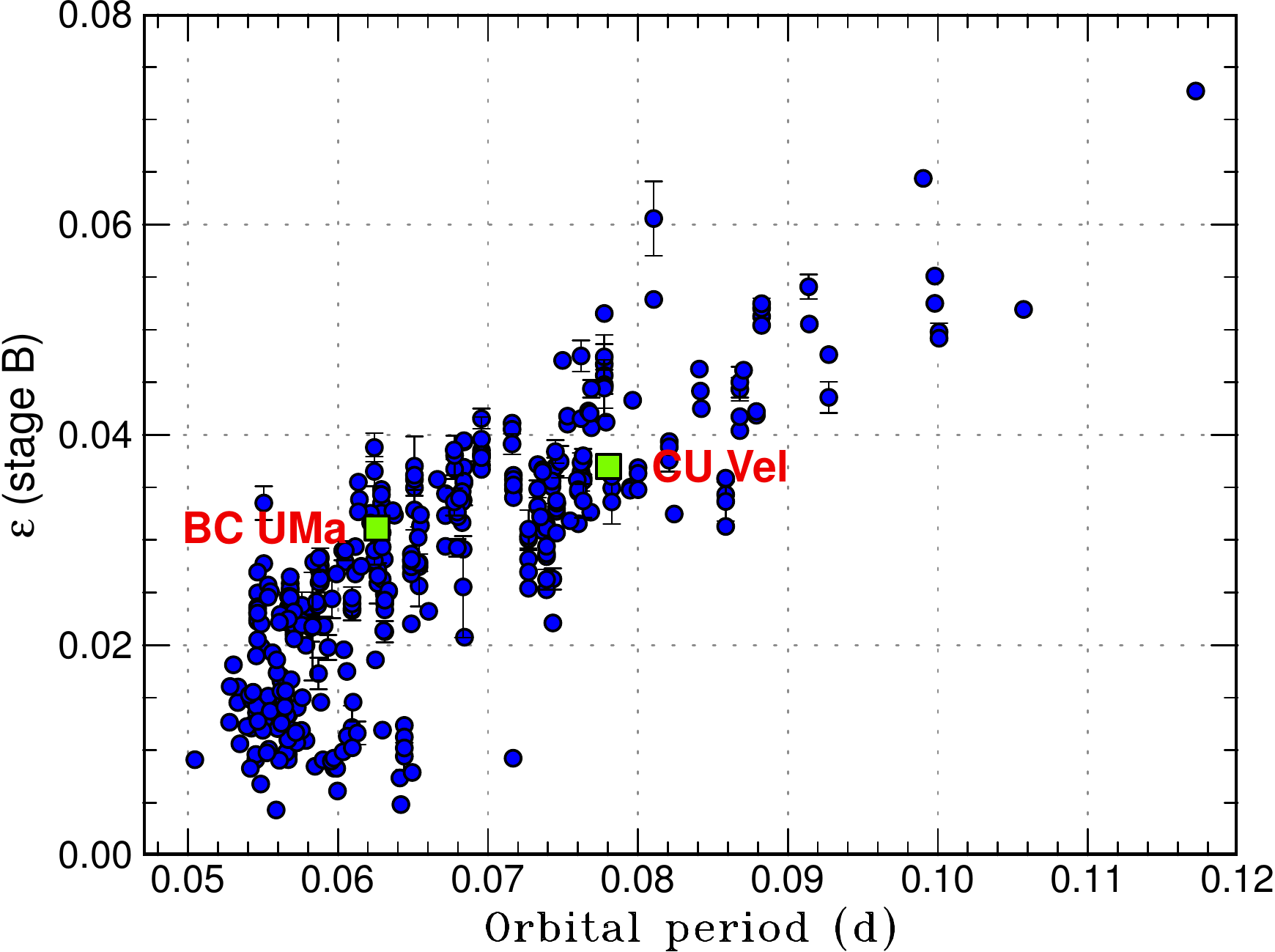}
  \end{center}
  \caption{Relation between $P_{\rm orb}$ and $\epsilon$ of
    stage B superhumps.  Although BC UMa and CU Vel have been
    reported have low-mass white dwarfs, they do not have
    unusually large $\epsilon$, suggesting that they have
    white dwarf masses typical for CVs.
  }
  \label{fig:stagebporb}
\end{figure*}

   \citet{pal21CVparam} published a compilation of
masses of white dwarfs ($M_{\rm WD}$) in 89 CVs.
While the mean mass $\langle M_{\rm WD}\rangle$ =
0.081$M_\odot$ was obtained, they identified five systems
with $M_{\rm WD} < 0.5M_\odot$ (table \ref{tab:lightWD}).
Since $M_{\rm WD}$ values for eclipsing SU UMa/WZ Sge stars
were known to be in a narrow region (table \ref{tab:meanWD}),
I examined whether the initial two SU UMa stars have
anomalously large $\epsilon$ (it is a pity that $P_{\rm orb}$
is now known for SDSS J100515.39$+$191108.0, superhump
observations were present).
For an 0.5$M_\odot$ white dwarf,
$\epsilon$ is expected to be 1.6 times larger than average
SU UMa stars with the same $P_{\rm orb}$.
The result is shown in figure \ref{fig:stagebporb}.
BC UMa and CU Vel do not have unusually large $\epsilon$,
and this figure suggests that they have white dwarf masses
typical for CVs.  The masses of white dwarfs in these
two objects were measured by
fitting the ultraviolet spectra obtained by
the Hubble Space Telescope (HST).  As judged from table 6
in \citet{pal21CVparam}, it looks like that this method is
responsible for a large number of small-mass white dwarfs.

   It might be worth noting that QZ Vir was reported to have
a very light ($< 0.4M_\odot$) white dwarf in the past
\citep{sha84tleo}, which has a completely normal $q$=0.108(3)
from the stage A superhump method.  This instance
also suggests that very large errors in the masses
of white dwarfs are expected by traditional methods.
\citet{zor11SDSSCVWDmass} wrote: ``Only 7$\pm$3\% of
the 104 CVs with available WD-mass estimates and errors
have $M_1 \leq 0.5M_\odot$ , and none of the systems in
the sub-sample of 32 with presumably more reliable mass
determinations. We therefore conclude that the fraction
of He-core WDs in the observed sample of CVs is $\leq$10\%''.
It may be that such low-mass white dwarfs are totally
missing in dwarf novae below the period gap considering
the absence of objects largely deviating from
the main distribution on the $P_{\rm orb}$-$\epsilon$ plane.

\begin{table*}
\caption{Very light ($< 0.5M_\odot$) white dwarfs in CVs.}\label{tab:lightWD}
\begin{center}
\begin{tabular}{lcc}
\hline\hline
Object             & $M_{\rm WD}$ & Method, comment \\
                   & ($M_\odot$) & \\
\hline
BC UMa             & 0.48$_{-0.09}^{+0.08}$ & Revised ultraviolet spectral fit to \citet{gan05bwsclbcumaswuma} \\
CU Vel             & 0.47$_{-0.05}^{+0.04}$ & Ultraviolet spectral fit \citep{pal17CVWD} \\
SDSS J100515.39$+$191108.0 & 0.44$_{-0.09}^{+0.15}$ & Ultraviolet spectral fit \citep{pal17CVWD} \\
V754 Lyr           & 0.46(2) & Modeling ellipsoidal light variations, \\
                   &         & low mass-transfer rate for $P_{\rm orb}$=0.3641~d
                               \citep{yu19v754lyr} \\
HY Eri             & 0.43$_{-0.07}^{+0.10}$ & \citet{bur99hyeri}, polar \\
\hline
\end{tabular}
\end{center}
\end{table*}

\begin{table*}
\caption{Masses of white dwarfs in CVs.}\label{tab:meanWD}
\begin{center}
\begin{tabular}{ccc}
\hline\hline
$\langle M_{\rm WD}\rangle$ below period gap & Intrinsic scatter & Reference \\
($M_\odot$) & ($M_\odot$) & \\
\hline
0.83$\pm$0.02 & 0.07 & \citet{sav11CVeclmass} \\
0.80         & 0.12 & \citet{zor11SDSSCVWDmass} \\
0.82$\pm$0.02 & 0.10 & \citet{mca19DNeclipse} \\
0.81          & $-$0.16/$+$0.17 & \citet{pal21CVparam} \\
\hline
\end{tabular}
\end{center}
\end{table*}

\section{Advantages and Disadvantages of Eclipse Modeling Method
and Stage A Superhump Method}

   I summarize the advantages and disadvantages of the two methods.

\begin{flushleft}
\large{\textbf{Eclipse modeling method}}
\end{flushleft}
\begin{flushleft}
Advantages:
\end{flushleft}

\begin{itemize}
\item Basically applicable to eclipsing systems regardless
of the dwarf nova type.
\item Considered to be the most reliable.
\item Both the mass and radius of the secondary can be measured
to directly determine the evolutionary state of the secondary.
\end{itemize}

\begin{flushleft}
Disadvantages:
\end{flushleft}

\begin{itemize}
\item Only applicable to eclipsing systems.  Inclinations
need to be high enough and cannot be directly applied to
grazing eclipsers.
\item Requires a large telescope and a high-speed imaging device.
\item Requires relatively bright ($\sim$20~mag) objects
in quiescence.
\item Systems without orbital humps are not good targets.
\item Affected by flickering in high mass-transfer systems.
\item Requires complicated modeling of the light curve and
an advanced numerical method such as (parallel-tempered) MCMC.
\end{itemize}

\begin{flushleft}
\large{\textbf{Stage A superhump method}}
\end{flushleft}
\begin{flushleft}
Advantages:
\end{flushleft}

\begin{itemize}
\item Can be applied to any inclination.
\item Small telescopes (20--60~cm) are usually sufficient.
\item Very high time resolutions are not needed.  Time resolutions
around 30~s are usually sufficient.
\item Relatively faint ($\sim$16~mag using the above equipment)
objects can be observed.
\item Less disturbed by flickering.
\item Period determination is numerically very easy.
\item A considerable fraction of WZ Sge stars show early superhumps
and orbital periods and $q$ values can be determined only by
outburst observations.
\item The sample is expected to grow steadily as new transients
are discovered.  The number now exceeded 100 and is 2.5 times
those determined by the eclipse modeling.
\end{itemize}

\begin{flushleft}
Disadvantages:
\end{flushleft}

\begin{itemize}
\item Only applicable to superhumping systems (SU UMa/WZ Sge stars).
Cannot be applied to SS Cyg stars.
\item Observations of the growing phase (stage A) of superhumps
are necessary.  Timely alert and intensive observations
(not always easy) during these superhumps are essential.
\item Systems with relative large $q$ have short durations
of stage A and the accuracy of the $q$ estimate is limited
by the short baseline of the observations.
\item Not very adequate for eclipsing systems.
\end{itemize}

\section*{Epilogue}

   Although I wrote this paper myself, this works has been
impossible without contributions of professional and amateur
astronomers who wish to catch and confirm the nature of suddenly
appearing new objects.  It is sometimes said that
birding (birdwatching) invokes our primeval hunting instincts,\footnote{
  See e.g. $<$http://www.birdwatching.com/birdingfaq.html$>$.
} and our activity with ``hunting superhumps'' may be invoked
by the same type of our primeval hunting instincts.
Since my childhood, I had been acquainted with media reports
telling a certain bird species first recorded in Japan or
spotted after absence of tens of years (these birds are
called ``vagrants'' or ``rarities'').  When I became
enchanted by the field of variable stars, particularly
``UG'' (=U Gem) stars (currently known as dwarf novae;
subclasses were not specified at that time),
I noticed the striking similarity between birding and
hunting rarely outbursting stars (recurrent novae and
rarely outbursting UG stars; WZ Sge was considered as
a recurrent nova at that time).  There were even
parallel terminology (jargon) both in birding and
UG hunting, at least in the Japanese language.
We were informing each other via telephone which stars
were in outburst; this is exactly what present-day birders
do with cell phones.  Some present-day observers, particularly
such as the renowned CV enthusiast Patrick Schmeer,
still know how I was devoted to hunting UG outbursts
(I have a vivid recollection of his international calls
when I was observing at the university observatory 
as a beginner professional: ``Hello, this is Patrick Schmeer
from Germany...", asking confirmation of his suspected
detection of outbursts).

   When I chose the carrier as a professional astronomer,
I switched my main hobby from variable star observations to
birding --- this was initially from curiosity about
what the world on the other side, connected by a similar
set of jargon, would look like.  I got, of course, enchanted
by the world of birds: not only because birding itself
was interesting (as interesting and deep as astronomy as
a hobby) but also due to the fascination of the birds
themselves.

   Looking back the history of studies of superhumps,
I feel that the present accomplishment, just like this paper,
can be regarded as one of the most remarkable achievements
of \textbf{citizen science} in astronomy --- it rivals, or may
even surpasses, the achievements by large telescopes.  
Many amateur and professional observers have devoted themselves
to detect outbursts and observe superhumps.  I greatly owe
this accomplishment to them (and of course, to the theoretician
Prof. Yoji Osaki for interpreting superoutbursts, superhumps
and the precession rate).
The existence of international groups played an important
role: our VSNET [and, of course, the Variable Star
Observers League in Japan (VSOLJ)]
had a main coverage in the East Asia and Europe,
while the Center for Backyard Astrophysics (CBA)\footnote{
  $<$https://cbastro.org$>$.
} had a main coverage in the North America and Europe, and
some southern hemisphere countries.
These two parties were the driving force of studies of
CVs in professional-amateur collaborations,
or citizen science in modern, more fashionable terminology.
I sometimes recall competitions between these groups
in the late 1990s and 2000s when both groups observed
SU UMa/WZ Sge stars.  I feel somewhat a pity
that CBA observers do not observe SU UMa/WZ Sge stars
as before.  Because of this, data above the skies
over the Americas currently tend to be lacking.

   Why the longitudinal coverage matters? --- This comes
from the nature of the transient objects.
Stage A superhumps in SU UMa stars usually last 1--2~d
in long-$P_{\rm orb}$ systems and a few days
in short-$P_{\rm orb}$ systems.
A good temporal coverage is necessary to obtain
a reliable period.  The durations are an order of a day,
not a few hours or several days, and this is
why a \textit{longitudinal} coverage is so important
since a given object can be observed only for several
hours (up to an entire night in extreme cases) 
from a given location on the globe.  This situation is
very different from objects like intermediate polars,
the current main targets of the CBA.
Intermediate polars do not require continuous coverage
and a longer baseline in time is more important, which
can be achieved by observations from a single location.
Considering these characteristics, dwarf novae are
ideal objects for uniting world-wide observers
particularly in the longitudinal direction.

   Looking at broader fields of variable stars,
the American Association of Variable Stars (AAVSO)\footnote{
  $<$https://www.aavso.org/$>$.
}
plays the central role, providing observational
database and the world most comprehensive
variable star catalog, the AAVSO Variable Star Index (VSX).
Researchers engaged in variable objects should
have already consulted the catalog or database provided
by the AAVSO.
The AAVSO is just celebrating the 110 years of
citizen science.\footnote{
  $<$https://www.aavso.org/110$>$.
}
Wouldn't it be one of the most renowned citizen science
programs with a long and distinguished history?

   Let's look at the other side of the world ---
I mean the world of birds.  The online database of
bird reports eBird\footnote{
  $<$https://ebird.org/$>$.
}
has a history of 20 years, although much less than the AAVSO,
provides the function similar to the AAVSO.
Birders among the readers of this paper should
certainly be aware of eBird and some may have already
submitted observations to it.  Again, the analogy
between variable star observing and birding is here.
I recommend readers who are not familiar with eBird
to visit there and explore the functions of eBird, and you
will find the similarity of these fields.
I even wonder if Sebasti\'an Otero, the chief supervisor
of VSX, is well
familiar with the other side of the world and his work
in astronomy is also inspired by activities in
the world of birding.
His page on the AAVSO website\footnote{
  $<$https://www.aavso.org/sebasti\'an-otero$>$.
} tells ``\textit{When he's not doing something related to
astronomy (and he's not sleeping) you may find him
in some natural reserve taking pictures of every bird
he is able to detect, because he is an active bird-watcher.
To find several weird species, he likes travelling all
around his country when he has time, so it seems
everything that is up in the sky deserves his attention!}''
I'm suspecting that he continuously gains a momentum
from his own birding activity or inspirations from activities
of birding organizations, including eBird, to his enthusiastic
tasks in variable stars.

   I have written that dwarf novae are ideal objects for
uniting world-wide observers particularly in the
\textit{longitudinal} direction.  What about the
\textit{latitudinal} direction?  We indeed collaborate
with southern hemisphere observers to study southern
object which we cannot see from the northern hemisphere,
but there are far better ``ambassadors of friendship''
in the world of birds --- migratory birds.
Many migratory birds annually move from the north to
the south and the reverse, and international collaboration
particularly in the \textit{latitudinal} direction
is indispensable for studies and conservation
of these birds.
I would propose to express that variable star observations
and studies of migratory birds are the best examples of
warp and woof (longitudinal and latitudinal threads)
of global scientific collaboration and friendship.

\begin{figure*}
  \begin{center}
    \includegraphics[height=6.5cm]{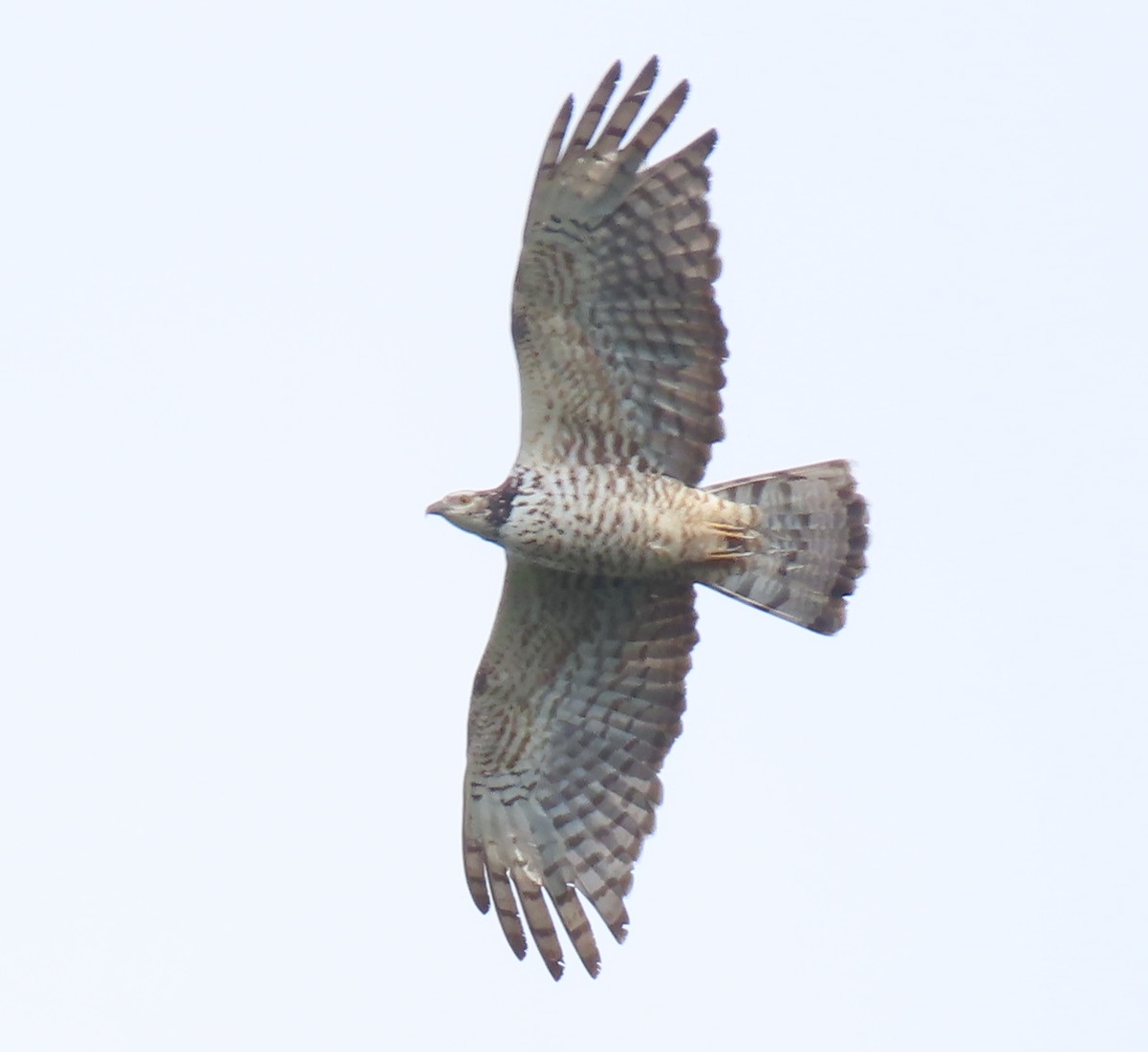}
    \includegraphics[height=6.5cm]{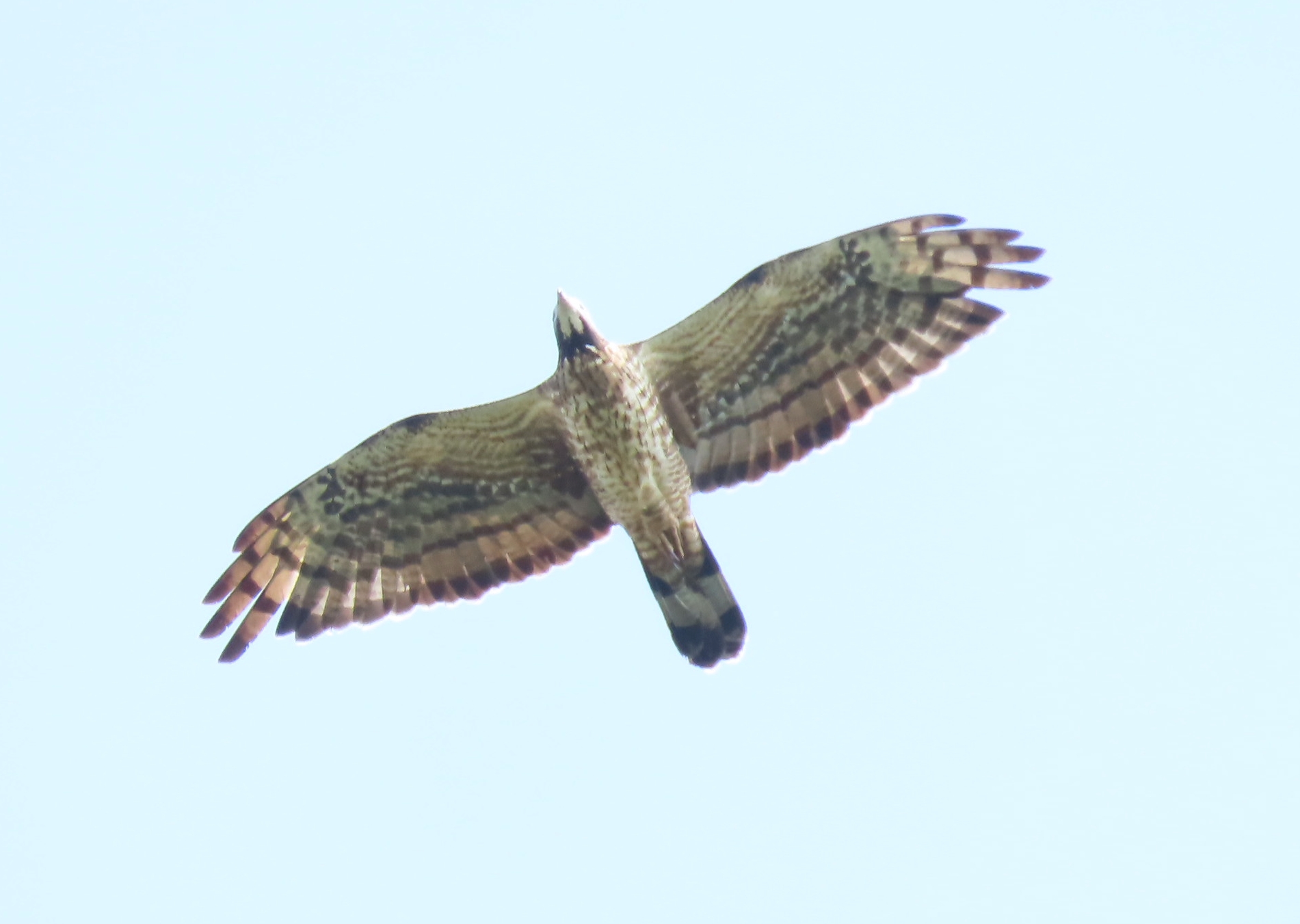} \\
    \includegraphics[width=16.35cm]{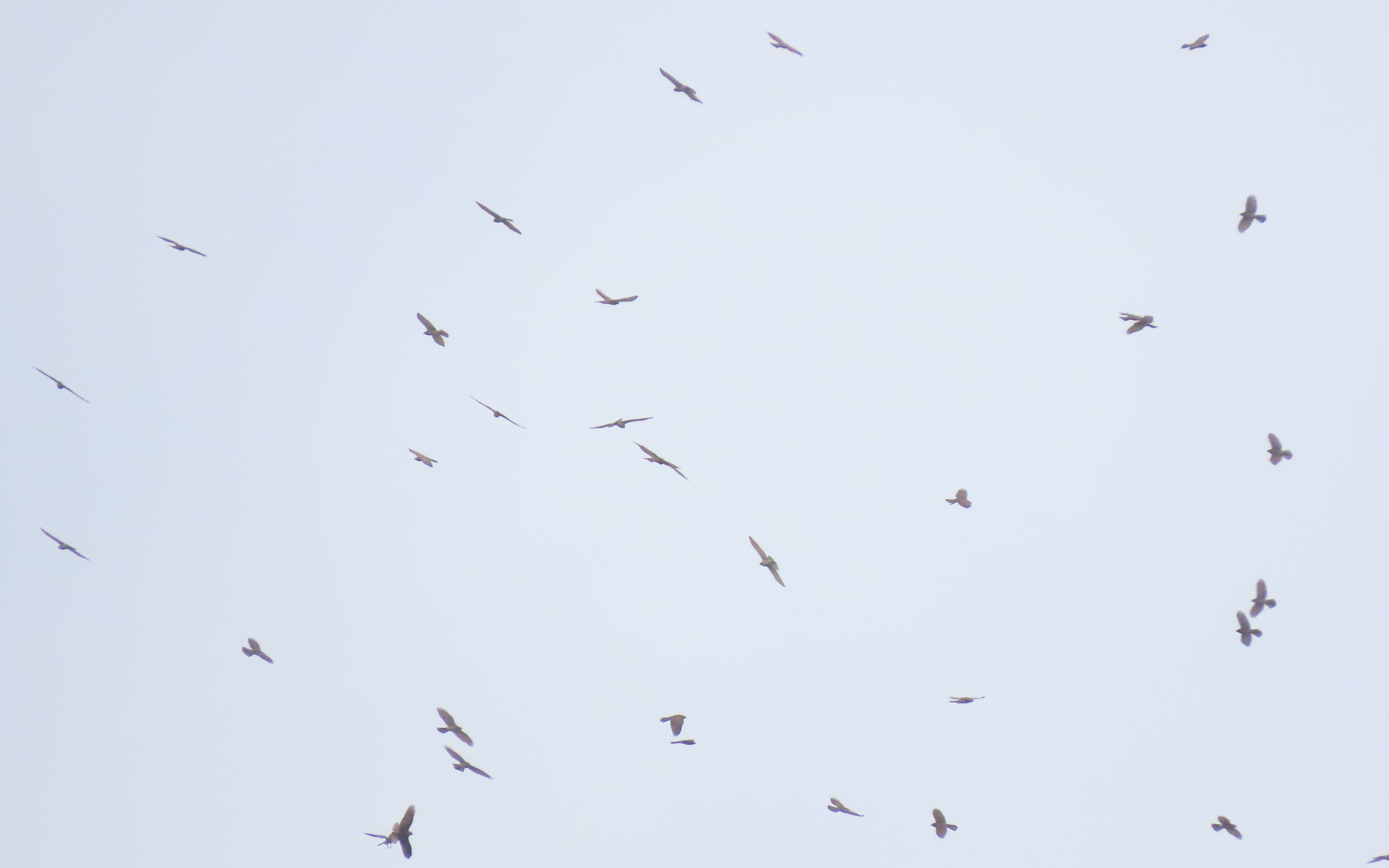}
  \end{center}
  \caption{Migratory raptors.
    (Upper:) Crested Honey Buzzards (\textit{Pernis ptilorhynchus})
    on migration.  Female (left) and male (right).
    (Lower:) Thermally soaring group (``kettle'') of
    Grey-faced Buzzards (\textit{Butastur indicus}) on migration
    (photos by author).}
  \label{fig:hawks}
\end{figure*}

   Just to name a few, shorebirds (waders) migrate from
the arctic tundra to Australia and many people along this
East Asian-Australasian flyway\footnote{
  East Asian-Australasian Flyway Partnership (EAAFP)
  $<$https://www.eaaflyway.net/$>$.
}
are engaged in observation and conservation.
Similar kinds of flyways are present in the Americas
(Pacific Americas flyway, Atlantic Americas flyway),
central Eurasia (Central Asian flyway), Europe-Africa
(East Atlantic Flyway) and others.\footnote{
   The names of flyways depend on the literature.
}
Some migratory birds are ``visible'' on migration.
One of the best examples is diurnal raptors
(eagles, hawks and allies)
[See e.g. \citet{bil06raptormigration,pan21raptormigration}].
There are hot spots of migration corridors
around the world, such as Batumi in Georgia famous
for the number of passing raptors \citep{ver11batumiraptor}.\footnote{
  $<$https://www.batumiraptorcount.org/$>$.
}
There was even a record of more than 200000 raptors
passing in a single day.  During the seasons of migration,
raptor watchers along the migratory route observe
the sky all the day just as meteor observers watch
the sky.  Among migratory raptors, I pick up
the Crested (or Oriental) Honey Buzzard
(\textit{Pernis ptilorhynchus}; figure \ref{fig:hawks}).
They breed in Russia, Japan, Korea and northeastern
part of China and migrate to the Southeast Asia to winter.
It has been demonstrated by satellite tracking
that this species encompasses all the East Asian
and Southeast Asian countries during migration!
\citep{hig05honeybuzzard}.
Enthusiastic watchers observe them along
the migration route, but the information is
still limited to a relatively small number of countries.
Birds sometimes travel beyond our imagination.
I was surprised to read the news that
a Crested Honey Buzzard was recorded first time in
the South African Republic in 2021
\citep{coh21honeybuzzard}.\footnote{
   See, for example,
   $<$http://www.alexaitkenhead.co.za/2021/02/crested-honey-buzzard-pernis.html$>$
(There is a photograph of people waiting for the bird to appear.
It was written: ``Official Confirmation registering the CHB
(=Crested Honey Buzzard)
as the first record for South Africa, making this raptor from
the orient the highest priority amongst the birding fraternity.
Saturday 6th February was our first opportunity to get onto
the view site and hopefully get a glimpse of this famous buzzard''.
There was also an interesting story at
$<$https://www.iol.co.za/capetimes/news/rare-sighting-of-a-crested-honey-buzzard-in-somerset-west-has-birders-abuzz-172fd2d5-7afd-42a2-a0e6-43821a5ca4b3$>$
from the Cape Times.
}
I had never imagined to make a collaboration with
a South African researcher about this oriental species,
although I had a number of joint papers with a variable star
observer in South Africa in the field of dwarf novae.
You can see from these pages and images
that the enthusiasm is the same between these two fields,
and feel the wonders of the nature.

   Scientific activities in astronomy and ornithology are
not totally independent.  Since migratory birds are known to
use the Sun, stars [this has been confirmed by experiments
using a planetarium as early as \citet{eml67celestialorientation,
eml67orientationmechanism}] and perhaps the Moon in addition to
geomagnetic cues for orientation.  Light-level geolocators
\citep{afa93geolocator}, tracking devices that use daylight to
estimate location, are now widely used in studying movements of
various types of birds \citep{cro05albatross,bri09geolocator}.
This technique is dependent on the solar ephemeris
(astronomical almanac).
I have even been invited to join a research program by
providing a handy code for calculating solar and lunar
ephemerides along the trajectory of moving birds
by using knowledge in astronomy \citep{hol22shiningcuckooo}.
Movements of birds at night in relation to the moonlight
(e.g. \cite{nor19lunarbirdmigration})
are likely a field in which astronomers can
contribute to ornithology.
We should not forget that counting migratory birds on
the disk of the Moon has long been used
(e.g. \cite{ver1897birdmoon}) --- many (particularly small)
birds migrate at night to avoid overheating in daytime
flight or to avoid predation.  The scene was also beautifully
described in Rachel Carson's famous book
``The Sense of Wonder'' (published posthumously:
\cite{car65senseofwonder}).  Migrating birds often issue
calls (nocturnal flight calls or NFC,
first described in \cite{bal52nocmig})
at night, probably communicating each other.
Recording of these calls (nocmig)\footnote{
  $<$https://nocmig.com/$>$.
}
is also an important tool to study bird migration
(e.g. \cite{eva99nocmig,lar02nocmig,fra04nocmig}).
The nights suitable for astronomical observations are
also suitable for migratory birds to move.
Won't someone working in suitable astronomical observatories
regularly make recordings of these nocturnal flight calls?

    Modern space science technologies also help biologists
to understand the behavior of moving animals and take part
in their conservation.  In particular, the ``Space for Birds''
project from the International Space Station (ISS) maps
the routes taken by seven endangered or
threatened bird species, highlighting along those routes
habitat changes caused mainly by human activities.\footnote{
  $<$https://www.nasa.gov/mission\_pages/station/research/news/amass-ceo$>$.
}
There is also the International Cooperation for Animal Research
Using Space (ICARUS) project using the ISS by collaboration
of the Max Planck Institute for Ornithology, Germany and
the Russian Space Agency Roscosmos.\footnote{
  $<$https://directory.eoportal.org/web/eoportal/satellite-missions/i/iss-icarus$>$.
}  This page explains: ``To achieve the ICARUS goals,
the following prerequisites have to be fulfilled: Global tag
coverage to record long distance migration patterns;
Simultaneous communication with a multitude of animal tags; 
Extremely low tag mass and size to allow tracking of small animals; 
Long, maintenance-free tag life in order to cover complete
migration cycles; and logging of the internal (physiological)
and the external (environmental) state of animals''.
The system went into operation in 2020 March, and the data
are stored in Movebank \citep{fie12movebank}.\footnote{
  $<$https://www.movebank.org/$>$.
}
The data are made publicly open and scientists around
the world can make their own analyses, just as we do
with observations of major astronomical observatories
and astronomical satellites.  Movebank also hosts other
sets of tracking data obtained by various researchers
worldwide.

   During the preparation of this paper, I heard the passing
of the renowned, and sometimes debated, biologist
Edward Osborne Wilson (1929-2021), who is not only renowned
for inventing the concept ``sociobiology'' in his book
``Sociobiology: The New Synthesis'' \citep{wil75sociobiology}
but also publishing ``The Diversity of Life''
\citep{wil92diversityoflife} in conjunction with
the 1992 United Nations Conference on Environment and Development,
Rio de Janeiro, Brazil (commonly known as the ``Earth Summit'').
This summit resulted ``Rio Convention'' including
``The Convention on Biological Diversity'' and
``The United Nations Framework Convention on Climate Change'',
both of which are the biggest issues facing our planet
and its inhabitants, including humans --- this will be
also true for extraterrestrial civilization, if it exists,
since any civilization must have evolved from
rich biological diversity.
The popularity of Wilson's book \citep{wil92diversityoflife}
was one of the motives to spread the concept of
biological diversity, or biodiversity.

   When I met the former book \citep{wil75sociobiology},
I was surprised to see that biologists, including Wilson
himself, applied Shannon entropy (information entropy,
\cite{sha48entropy}) in describing and discussing
the evolution of communication in biology [in his chapter 8;
biologists wanted to explain why the number of types of
signals is highly conserved (10--40) between vastly different
species; this chapter is a very good introduction to
the theory of communication in biology; if the book is
in your library, it is certainly worth looking at]
so early in the history.
Both \citet{hal54insectcommunication} and
\citet{wil62antcommunication} used the expression of entropy
for continuous variables \citep{sha49entropycommunication}
to describe animal behavior.  \citet{haz65communication}
measured transition probabilities of Markov chains
dealing with three types of behavior of crabs and estimated
the information entropy associated with communication.
We astronomers tend to think (sorry if it is my
misunderstanding or prejudice!) that biologists are less
mathematically inclined, but this is not always true.
The maximum entropy method (MEM) is the most frequently used
form of Shannon entropy in astronomy.  By searching using
ADS, the first applications of MEM in astronomy appears
to be \citet{ric73WDMEM,ric74WDMEM} [These papers cited
applications to geophysics: \citet{ulr72MEM}].
The first application of MEM to inversion problems,
which we meet more frequently now, appeared
in \citet{bry80m87jetMEM}.  Compared to them, you can see
how early the applications to biology were.

   E. O. Wilson left ``Advice to young scientists''
in his TED talk in 2012.\footnote{
  $<$https://www.ted.com/talks/e\_o\_wilson\_advice\_to\_a\_young\_scientist$>$.
}
Although the main audience of this talk was apparently young
biologists or medical students and the talk as a whole
may not be directly applicable to young astronomers of
physicists, he stated: ``\textit{In time, all of science will come
to be a continuum of description, an explanation of
networks, of principles and laws. That's why you need
not just be training in one specialty, but also acquire
breadth in other fields, related to and even distant
from your own initial choice. 
Keep your eyes lifted and your head turning. The search
for knowledge is in our genes. It was put there by our
distant ancestors who spread across the world, and it's
never going to be quenched}''.  This part of the talk
probably referred to his concept of ``Consilience''
\citep{wil98consilience}, although humans have not
yet reached a point to ``unite the sciences and might
in the future unite them with the humanities''
as proposed by Wilson.  This talk, however, conveys
an important message for us, too.
Keeping the eyes open to different
fields of science is undoubtedly important.
For example, by applying the existing methods to other fields
of science, one may be able to obtain a deeper insight
and a positive feedback to the original field,
just as I attempted in \citet{kat21lasso2}.
Such multidisciplinary applications
are always appreciated for development of science.
As I have shown, two fields of science, variable star observing
(astronomy) and birding (ornithology), which are usually
considered to be distant in terms of science,
are two sides of the same coin in human activity invoked by
our primeval hunting instincts.  It would be easy to
understand the fun of the other side from a different side
and join or respect the other activity.
More importantly, these two activities share the same character
in that they are essentially international collaborations
and use the similar types of information such as public
databases.  These activities have a common aspect that
they contribute to uniting people worldwide via exchanges
of observations.  If experience, knowledge and methods 
of the two fields are mutually exchanged, these fields
would be expected to be even more powerful.

   Astronomers can contribute to ornithology
by introducing techniques (including mathematical modeling
or computer science) or ideas used in astronomy
(astronomy is a bit ahead of ornithology in the use
of the literature database ADS, the preprint server arXiv
and open databases, and I hope that ornithological community
assimilates the usefulness of these services and introduces
some of functions in the astronomical services
into their activity).
Although the migration routes of
the Crested Honey Buzzard are very different between spring
and fall migrations \citep{hig05honeybuzzard,yam08honeybuzzard},
the reason of the complexity of the routes is not
yet well understood.
Biologists have considered the weather (particularly
assisting winds) during migration and foraging ecology
at stopover sites
\citep{yam12honeybuzzardcrossing,mar15honeybuzzardhabitat,
nou17honeybuzzardclimate,sug19honeybuzzard} to be main reasons.
\citet{ago07honeybuzzard}, however, claimed that food diversity
and abundance along the migration route should not be of
critical importance and suggested that there might not yet
have been sufficient time to evolve the more direct route
through the southwest islands.  Astronomers may provide ideas
to solve the issue by using mathematical modeling or
by the help of data-driven science.
Although the tracking data of the Crested Honey Buzzard are
unfortunately not yet available on Movebank, there are
public data for the European Honey Buzzard
(\textit{Pernis apivorus}), the species sister to
the Crested Honey Buzzard, and other migratory raptors.
They can be used for modeling the migration routes.
This is only an example, and I feel that astronomers can
contribute to ornithology in various aspects using public data.

   In order to make an actual contribution, basic knowledge
and experience are required to some extent.
Birding would be one of ideal tools to start with to become
familiar with biology (if one is not yet), since we can directly
learn from our winged neighbors.
It may be needless to say, but I am tempted to add
a bit (a lot?) about ``birdbrain''
for readers who are not familiar
with modern biological development.
``Birdbrain'' literally means a stupid person,
or a person lacking intelligence.
It was indeed written in old textbooks
that birds lack the part of the brain (cerebral cortex),
which is present in mammals, and the prefrontal cortex
in the cerebral cortex is known to be particularly
relevant for intelligence.
I indeed learned that behavior of birds is simply
a complex collection of reflexes and instinctual behavior
governed by the basal ganglia (old part of the brain).
It has become, however, apparent that hyperpallium
(in modern terminology) in birds is equivalent to
the mammalian cerebral cortex and a consortium of
neuroscientists has proposed renaming of the structures of
the bird brain \citep{jar05birdbrain,her20birdpallium},
which can be directly compared to the mammalian brain
(following the nomenclature common to birds and mammals,
the mammalian neocortex is also called neopallium).
It has been even shown that birds have numbers of neurons
in the forebrain (advanced part of the brain) comparable to
the primates \citep{olk16birdneuron}.  These results have
already been reflected on the wikipedia page\footnote{
  $<$https://en.wikipedia.org/wiki/List\_of\_animals\_by\_number\_of\_neurons$>$.
}
and the comparison is amazing.
The Rook (\textit{Corvus frugilegus}), a relatively small
species of a crow, has a number of neurons comparable to
a Beagle dog and exceeds that of
the Lion (\textit{Panthera leo})!
Some bird species have cognitive ability rivalling
great apes [such as tool use, mirror recognition, plan for
future needs and vocal learning;
for a review see \citet{olk16birdneuron}].
It has been shown that the working memory has the same
neuronal mechanism between crows and monkeys
\citep{hah21crowmonkey}.  The nidopallium caudolaterale
in birds is considered to be equivalent to
the prefrontal cortex in mammals.  It is also known to
work when pigeons discriminate the abstract features
of paintings \citep{and20pigeonnidopallium}.
Avian nidopallium caudolaterale and mammalian 
prefrontal cortex are considered to be an example of
convergent evolution and it has been suggested that
there may be limited degrees of freedom in developing
intelligence \citep{gun05nidopallium}.
The intelligence of birds is probably a part of
the reason why we understand birds easily and
birds attract many people.
I can list at least three highly intelligent groups of birds:
parrots (very well-known), crows (annoy and
sometimes amuse us\footnote{
   See e.g. ``Crowboarding: Russian roof-surfin' bird caught on tape'':
   $<$https://www.youtube.com/watch?v=3dWw9GLcOeA$>$.
}
everywhere) and the Striated Caracara
(\textit{Phalcoboenus australis}).
A Grey Parrot (\textit{Psittacus erithacus}) named
Alex (1976-2007)\footnote{
  $<$https://en.wikipedia.org/wiki/Alex\_(parrot)$>$.
}
could even make elementary vocal conversations with humans.
The New Caledonian Crow (\textit{Corvus moneduloides})\footnote{
  $<$https://en.wikipedia.org/wiki/New\_Caledonian\_crow$>$.
}
could use, store and even manufacture tools
\citep{wei02newcaledoniancrow,bay18newcaledoniancrow}.
The Striated Caracara is related to falcons and would be
less known (even to ornithologists),
but there are a number of YouTube videos
of behavior in the wild and laboratory experiments.
Just have a look at ``Flying Devils''
provided by National Geographic (2007).\footnote{
  $<$https://www.youtube.com/watch?v=Y7qcNiJTfVU$>$.
}
There is even a book ``A Most Remarkable Creature -- The Hidden
Life and Epic Journey of the World's Smartest
Birds of Prey'' \citep{mei21caracara}.
The important point is that intelligence evolved multiple
times in different lineages (this would be also true for
mammals considering intelligent dolphins or elephants).
We tend to think that intelligence evolved on a single path
within the primates and we humans are on the top of it,
but this picture may be an oversimplification of
the evolutionary process of intelligence.
The diversification time between birds and mammals
is estimated to be 297--326 million years.\footnote{
  $<$http://www.timetree.org/$>$.
}
Considering that all the three groups of very intelligent
birds belong the modern clade Eufalconimorphae
(diversification time of 73--87 million years from
other clades of birds), the appearance of the clades
hosting intelligent species independently from mammals
required $\sim$230 million years
since the diversification between birds and mammals.
This is nearly 40\% of time of 600 million years of
evolution of multicellular organisms \citep{che14multicellular}.
This value could be a measure to consider the evolutionary
time-scales of intelligence.
This lesson learned from very intelligent birds
may provide previously neglected insights into the life
helpful for dreaming of or searching for extraterrestrial
intelligence.  Yes, learning about birds
in modern perspective could contribute to understanding
of ourselves and possibly (distantly?) to astrobiology.

   We can also learn from the field of ornithology such as
eBird and other databases like xeno-canto\footnote{
  $<$http://www.xeno-canto.org/$>$.
}, which deals with vocalizations of birds.
The xeno-canto service is tolerant to multiple languages,
and data can be posted or discussed sometimes using
non-English languages (even in languages distant from
English, such as Chinese and Russian).
Such an aspect may be missing in USA-based systems
like the AAVSO and eBird, and xeno-canto could be
a model for multilingual international citizen science
programs.
We can also learn more broadly from ecology or biology. 
There are methods in ecology with which astronomers
are not usually familiar.  For example, the open source
software ``MaxEnt''\footnote{
  $<$https://biodiversityinformatics.amnh.org/open\_source/maxent/$>$.
}
\citep{phi04maxent,phi06maxent,eli11maxent} is widely used
in landscape ecology for modeling species' distributions
and selection of explanatory variables (e.g. choosing factors
which determine the presence of a certain species and
a prediction of the distribution of a certain species when
observations are fragmentary).  This method is an extension of
the well-known logistic regression.
We might be urged to apply Least Absolute Shrinkage
and Selection Operator (Lasso: \cite{Lasso}) to such a class
of problems, but MaxEnt can handle the data in which
non-detection data points are largely missing and is likely
a better solution under such circumstances.  Such a method
may inspire applications in astronomy under similar
circumstances.

   Returning to E. O. Wilson, there is a concept of
the Biophilia hypothesis, which suggests
that humans possess an innate tendency to seek connections
with nature and other forms of life (from wikipedia).
\citet{wil84biophilia} defined it in his book ``Biophilia'' as
``\textit{the urge to affiliate with other forms of life}''.
The current popularity of astrobiology and related fields
may be a manifestation of the ``Biophilia gene''
(Wilson might have already noticed this, but he would
have been happy to know this idea if he were still alive).
In some future, humans may build space colonies and
live within them.  This is quite understandable since
``\textit{the search for knowledge is in our genes. It was put
there by our distant ancestors who spread across the world,
and it's never going to be quenched}'' (Wilson) and
spreading outside the Earth would be a natural outcome
of expression of these genes.
In such space colonies, we can eventually test
the Biophilia hypothesis.  In isolated artificial environments,
people with strong expression of the Biophilia gene
may miss the lack of biodiversity, or may miss the absence
of migratory birds seasonally passing over or visiting us.
I hope that our distant descendants at that time still
have a healthy, but fragile, living planet to return
and can enjoy biodiversity and sometimes complain of
the weather when observing stars or waiting for meteors.

\newpage

\begin{abstract}
\xxinput{abst.inc}
\end{abstract}

\textbf{Key words:}
          accretion, accretion disks
          --- astronomical data bases
          --- catalogs
          --- methods: data analysis
          --- stars: binaries: eclipsing
          --- stars: dwarf novae
          --- stars: novae, cataclysmic variables
          --- stars: variables
          --- SU UMa stars
          --- WZ Sge stars
          --- superhumps
          --- mass ratios
          --- period bouncers
          --- brown dwarfs
          --- citizen science
          --- methods in biology
          --- bird migration
          --- raptor migration
          --- avian biology
          --- evolution of intelligence

\section*{Acknowledgements}

This work was supported by JSPS KAKENHI Grant Number 21K03616.
This research has made use of NASA's Astrophysics Data System.
I am deeply indebted to world-wide observers to study
superhumps.  Complete lists of collaborators have been given
in our Pdot papers.

\section*{List of objects in this paper}

\xxinput{objlist.inc}

\section*{References}

  I provide two forms of the references section (for ADS
and as published) so that the references can be easily
incorporated into ADS.

\renewcommand\refname{\textbf{References (for ADS)}}

\newcommand{\noop}[1]{}\newcommand{\hyphalt}{-}

\xxinput{shper2aph.bbl}

\renewcommand\refname{\textbf{References (as published)}}

\xxinput{shper2.bbl.vsolj}


\end{document}